\documentclass[pra, notitlepage, twocolumn, nofootinbib, floatfix, superscriptaddress, english, 10pt]{revtex4-1}
\usepackage{soul}
\usepackage[cm]{fullpage}
\usepackage[T1]{fontenc}
\usepackage[utf8]{inputenc}

\usepackage{amssymb}
\usepackage{amsmath}
\usepackage{amsthm}
\makeatletter
\def\amsbb{\use@mathgroup \M@U \symAMSb}
\makeatother
\usepackage[mathscr]{euscript}
\usepackage{mathtools}
\usepackage{verbatim}
\usepackage{bbold}
\usepackage{ifthen}
\usepackage{array}
\usepackage{collectbox}
\usepackage{mdframed}

\mdfdefinestyle{MyFrame}{%
    linecolor=black,
    outerlinewidth=8pt,
    roundcorner=20pt,
    innertopmargin=\baselineskip,
    innerbottommargin=\baselineskip,
    innerrightmargin=20pt,
    innerleftmargin=20pt,
    backgroundcolor=gray!10!white}

\usepackage{multirow}
\usepackage{placeins}
\usepackage[usenames,dvipsnames,table]{xcolor}
\definecolor{darkred}{RGB}{200, 0, 0}
\definecolor{darkgreen}{RGB}{0, 100, 0}
\definecolor{darkblue}{RGB}{0, 0, 200}

\definecolor{grun}{HTML}{00CED1}
\definecolor{gelb}{HTML}{FFEB00}
\definecolor{magenta}{HTML}{E36EFF}
\definecolor{cite}{HTML}{4C8383}
\definecolor{sec}{HTML}{69486f}
\usepackage{paralist}
\usepackage{textcomp} 
\usepackage{gensymb}
\usepackage{aliascnt}
\newtheoremstyle{lemeq}
{}
{}
{}
{}
{\bfseries}
{:}
{\parindent}
{\thmname{#1} \thmnumber{\normalfont{#2}} \thmnote{\normalfont#3}}
\theoremstyle{lemeq}

\newaliascnt{axiom}{theo}
\def\erw #1{\langle #1\rangle}
\def\nnnl{\nonumber\\}%

\def\id{\mathbf{1}}

\def\p{\varphi}%
\def\e{\varepsilon}%
\def\lt{(1-\lambda)}%

\def\und{\text{ and }}%

\newif\ifbem
\bemtrue

\definecolor{grun}{HTML}{00CED1}
\definecolor{gelb}{HTML}{FFEB00}
\definecolor{magenta}{HTML}{E36EFF}
\usepackage{pdfpages}
\usepackage{afterpage}
\usepackage{hyperref}
\hypersetup{colorlinks, linkbordercolor = {white}, breaklinks, linkcolor=sec, urlcolor=darkblue, citecolor=cite}

\makeatletter
\AtBeginDocument{\let\LS@rot\@undefined}
\makeatother

\begin{document}

\title{Device-Independent Quantum Key Distribution with Random Key Basis}

\author{Ren\'{e} Schwonnek}
\thanks{These authors contributed equally to this work.}
\affiliation{Department of Electrical \& Computer Engineering, National University of Singapore, Singapore}

\author{Koon Tong Goh}
\thanks{These authors contributed equally to this work.}
\affiliation{Department of Electrical \& Computer Engineering, National University of Singapore, Singapore}

\author{Ignatius W. Primaatmaja}
\affiliation{Centre for Quantum Technologies, National University of Singapore, Singapore}

\author{Ernest Y.-Z. Tan}
\affiliation{Institute for Theoretical Physics, ETH Z\"{u}rich, Switzerland}

\author{Ramona Wolf}
\affiliation{Institut f\"{u}r Theoretische Physik, Leibniz Universit\"{a}t Hannover, Germany}

\author{Valerio Scarani}
\affiliation{Centre for Quantum Technologies, National University of Singapore, Singapore}
\affiliation{Department of Physics, National University of Singapore, Singapore}

\author{Charles C.-W. Lim}
\email{charles.lim@nus.edu.sg}
\affiliation{Department of Electrical \& Computer Engineering, National University of Singapore, Singapore}
\affiliation{Centre for Quantum Technologies, National University of Singapore, Singapore} 

\begin{abstract}
\textbf{\abstractname}.
Device-independent quantum key distribution (DIQKD) is the art of using untrusted devices to distribute secret keys in an insecure network. It thus represents the ultimate form of cryptography, offering not only information-theoretic security against channel attacks, but also against attacks exploiting implementation loopholes. In recent years, much progress has been made towards realising the first DIQKD experiments, but current proposals are just out of reach of today’s loophole-free Bell experiments. Here, we significantly narrow the gap between the theory and practice of DIQKD with a simple variant of the original protocol based on the celebrated Clauser-Horne-Shimony-Holt (CHSH) Bell inequality. By using two randomly chosen key generating bases instead of one, we show that our protocol significantly improves over the original DIQKD protocol, enabling positive keys in the high noise regime for the first time. We also compute the finite-key security of the protocol for general attacks, showing that approximately $10^8$--$10^{10}$ measurement rounds are needed to achieve positive rates using state-of-the-art experimental parameters. Our proposed DIQKD protocol thus represents a highly promising path towards the first realisation of DIQKD in practice.

\end{abstract}

\maketitle

\section*{Introduction}
\begin{figure}[t]
    \centering
     \includegraphics[width=1\linewidth]{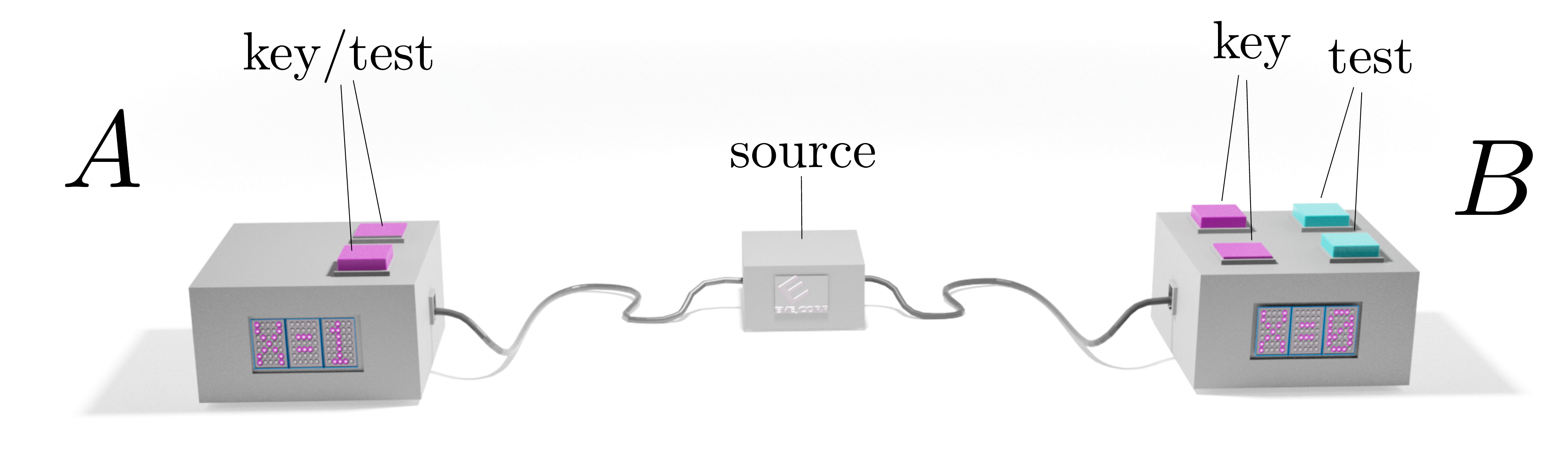}
     \caption{\textbf{Robust DIQKD:} Alice and Bob use uncharacterised devices to perform measurements on a quantum state that is created by a source that is potentially controlled by an adversary (Eve). In the proposed protocol, Alice has two possible inputs (measurement settings, magenta buttons) which are used for key generation and for running the CHSH Bell test, and Bob has four possible inputs grouped into two sets (magenta/cyan buttons): the magenta buttons are used for key generation while the cyan buttons are used for running the CHSH Bell test.} 
  \label{fig:boxes}
 \end{figure}

The basic task of DIQKD~\cite{mayers1998quantum,pironio,acin2007device,barrett2005no,reichardt2013classical} is to distribute a pair of identical secret keys between two users, called Alice and Bob, who are embedded in an untrusted network. To help them in their task, Alice and Bob are each given a measurement device, which they use to perform random measurements on a sequence of entangled systems provided by an adversary called Eve (see Fig.~\ref{fig:boxes}). The main advantage of DIQKD is that the measurement devices need not be characterised---Alice and Bob only need to verify that the input-output statistics of the devices violate a CHSH Bell inequality~\cite{clauser1969proposed,Bell1964}. As such, DIQKD represents the pinnacle of cryptography in terms of the number of assumptions required. More specifically, it only asks that (1) the users each hold a trusted source of local randomness, (2) their laboratories are well isolated, (3) they use trusted algorithms for processing their measurement data, (4) if the devices are reused for multiple instances of the protocol, the outputs in later instances do not leak information about earlier outputs, (5) they possess sufficient pre-shared keys to implement information-theoretically secure authenticated (public) channels, and that (6) quantum theory is correct. Given these basic assumptions (which are in fact standard assumptions in cryptography), one can then show that DIQKD is information-theoretically secure~\cite{miller2016robust,vazirani2019fully,arnon2019simple}. We note that assumption (4) is needed to address issues with protocol composition~\cite{PR21} and memory attacks~\cite{barrett2013memory}, because information-theoretic security may be violated if the protocol's public communication leaks some information about the private data from earlier instances.

The practical implementation of DIQKD, however, remains a major scientific challenge. This is mainly due to the need to have extremely good channel parameters (i.e., high Bell violation and low bit error rate), which in practice requires ultra-low-noise setups with very high detection efficiencies; though in recent years the gap between the theory and practice has been significantly reduced owing to more powerful proof techniques~~\cite{vazirani2014robust,vazirani2019fully,arnon2019simple} and the demonstrations of loophole-free Bell experiments~\cite{Hensen2015,Giustina2015,Shalm2015,Rosenfeld2017}. The present gap is best illustrated by Murta et al.~\cite{Murta2018}, whose feasibility study showed that current loophole-free Bell experiments are just short of generating positive key rates assuming the original DIQKD protocol~\cite{pironio,acin2007device} (see the dashed line in Fig.~\ref{fig:rates}).
\begin{figure*}[t]
        \includegraphics[width=18cm]{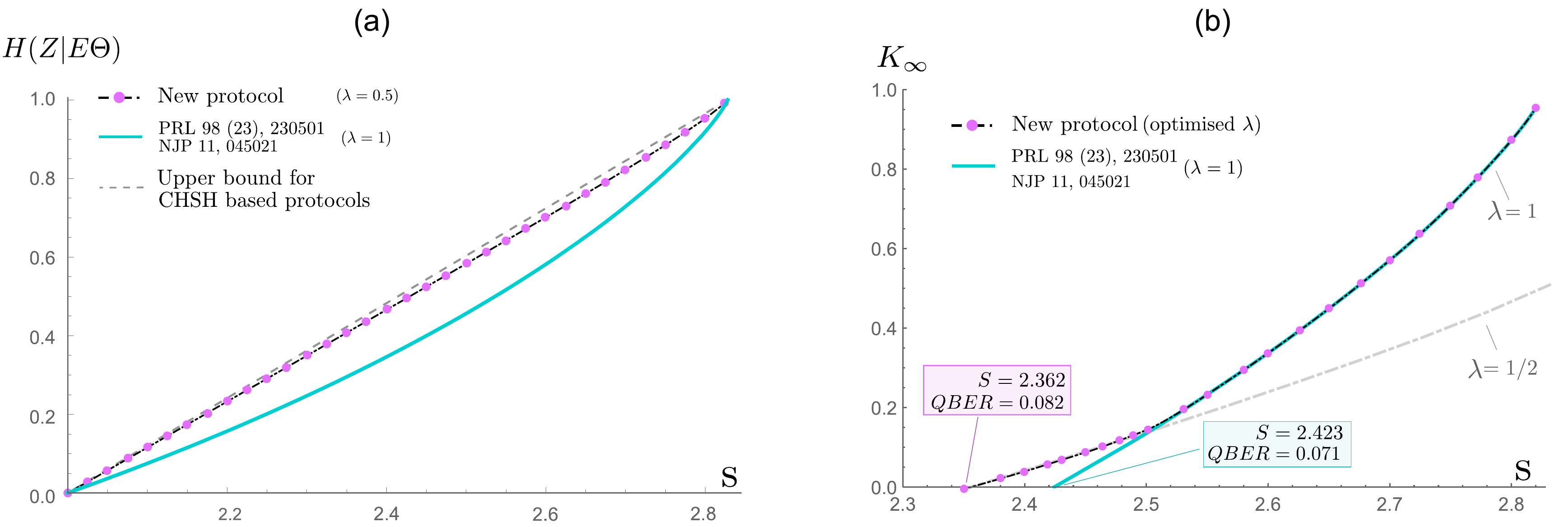}
        
         \caption{\textbf{Secret key rate and uncertainty of Eve:}
(a) Assuming the validity of quantum theory and a given CHSH value $S>2$, we show  that our new protocol can establish and certify drastically more uncertainty $H(Z|E\Theta)$ (close to the upper physical limit) than the best approach known before. (b) We consider a noise model (depolarising noise)~\cite{pironio,acin2007device} that only depends on the CHSH value $S$. Now a larger amount of noise can be tolerated in order to establish a positive key rate $K_\infty$. In detail we can decrease the critical CHSH value from $2.423$ to $2.362$. This corresponds to an increase of the critical bit-error-rate from $0.071$ to $0.082$, which brings a practical implementation of DIQKD into the reach of existing experiments.
    }
    \label{fig:ratesentro}
\end{figure*}
To improve the robustness of DIQKD, researchers have taken several approaches, from heralding-type solutions~\cite{gisin2010proposal,Moroder2011,Kolodynski2020}, local precertification \cite{ac1,ac2}, local Bell tests~\cite{limdevice}, to two-way classical protocols~\cite{tan2020advantage}. However, none of these proposals are truly practical, for they are either more complex in implementation or provide only very little improvements in the channel parameters. 

Here, we show that a simple variant of the original DIQKD protocol is enough to obtain significant improvements in the channel parameters. 

\section*{Results and Discussion}

To start with, we note that in the original protocol introduced by Ac\'{i}n et al.~\cite{acin2007device}, the key generating basis is predetermined and known to Eve. For concreteness, let Alice's and Bob's measurement settings be denoted by $X\in\{0,1\}$ and $Y\in\{0,1,2\}$, respectively, and let the corresponding outcomes be denoted by $A_X\in\{0,1\}$ and $B_Y\in\{0,1\}$. The secret key is derived from the events in which Alice and Bob choose $X=0$ and $Y=0$, respectively. The remaining measurement combinations are then used for determining the CHSH violation. Our DIQKD proposal is essentially the same as the original protocol, except that we introduce an additional measurement setting for Bob and now generate the secret key from both of Alice's measurements. This additional setting is needed so that Bob has a measurement that is aligned with Alice's additional key generating basis to obtain correlated outcomes (like in the case of the original protocol). Hence in our proposal, the key generation events are those where Alice and Bob choose $X=Y=0$ and $X=Y=1$. Below, we describe the proposal in detail.\newline

\begin{mdframed}[style=MyFrame]

Box 1 - \textbf{Proposed Device-Independent Quantum Key Distribution Protocol.} \newline

\noindent\textbf{1.~Measurements:}~This step is carried out in $N$ rounds, where $N$ is assumed to be asymptotically large. In each measurement round, Alice's and Bob's inputs, denoted by $X\in\{0,1\}$ and $Y\in\{0,1,2,3\}$, respectively, are drawn according to the following probability distributions: $P(X=0)=p$, $P(X=1)=1-p$, $P(Y=0)=q p$, $P(Y=1)=q(1-p)$ and $P(Y=2)=P(Y=3)=(1-q)/2$, where $0\leq p, q\leq 1$. Once Alice and Bob enter their inputs into their respective devices, they each obtain a measurement outcome, which we denote by $A_X\in \{0,1\}$ and $B_Y\in \{0,1\}$ respectively.\newline

\noindent\textbf{2.~Sifting:} Alice and Bob announce their measurement inputs over an authenticated public channel. This allows them to identify two common subsets of their measurement data: a pair of raw keys~\cite{scarani2009security} of size $\sim q( p^2+(1-p)^2)N$ (corresponding to $Y\in\{0,1\}$ and $X=Y$) and a pair of parameter estimation data of size $\sim(1-q)N$ (corresponding to $Y\in\{2,3\}$). Alice and Bob discard the remaining measurement data.\newline

\noindent\textbf{3.~Parameter estimation:} Alice and Bob publicly reveal their measurement outcomes from the parameter estimation data set and compute the underlying CHSH value:
\begin{equation*} S=\max\{2, C_{12}-C_{02}-C_{03}-C_{13}\},\end{equation*}where $C_{XY}=P(A_X =B_Y|X,Y)-P(A_X\neq B_Y|X,Y)$ is the correlation function of $X,Y$. Alice and Bob proceed to the next step if $S>S_{\rm{tol}}$, where $S_{\rm{tol}}$ is a predefined threshold value. Otherwise, they abort the protocol.\newline

\noindent\textbf{4.~One-way error correction and verification:} In the first part, Alice computes a syndrome based on her raw key (denoted by $\mathbf{L}$) and sends it to Bob via the public channel, who then uses the syndrome and his raw key to recover Alice's key. In the second part, they perform an error verification by comparing the hash values of their raw keys. Alice and Bob proceed to privacy amplification if the hash values are identical, otherwise they abort the protocol.\newline
    
\noindent\textbf{5.~Privacy amplification:} Alice and Bob perform privacy amplification to remove Eve's information about Alice's raw key. Once this is completed, Alice and Bob are left with a pair of identical secret keys.\newline

\end{mdframed}

In the parameter estimation step of the protocol, note that when the inputs are not uniformly distributed i.e. $p\neq1/2$, the CHSH value is to be computed in terms of the conditional probabilities $P(A_X,B_Y|X,Y)$ rather than the unconditioned probabilities $P(A_X,B_Y,X,Y)$ directly.
We remark that this does not introduce a measurement-dependence~\cite{hall11} security loophole, because the choice of inputs is still independent of the state.

It is well known that incompatible measurements are necessary for the violation of a Bell inequality and that such measurements are not jointly measurable and hence cannot admit a joint distribution~\cite{Fine82,busch,Wolf2009}. The intuition behind our proposal roughly follows along this line and exploits two related facts: (1) the key generation measurements of Alice must be incompatible for $S>2$ and (2) Eve has to guess the secret key from two randomly chosen incompatible measurements. 

When the secret key is only generated from a single measurement, like in the original DIQKD protocol, Eve's attacks are basically limited only by the observed CHSH violation and thus the monogamy of entanglement~\cite{HorodeckiRMP}. Eve, however, knows which measurement is used for key generation and hence can optimise her attack accordingly. 
On the other hand, if the secret key is generated from a random choice of two possible measurements, Eve faces an additional difficulty. Namely, in order to achieve a CHSH violation, the two measurements cannot be the same, and it is known~\cite{pironio} that for CHSH-based protocols, different measurements can give Eve different amounts of side-information; note that this is not the case for BB84 and six-state QKD protocols. Therefore, at least one of the measurements will not be the one that maximises Eve's side-information, giving an advantage over protocols based only on one key-generating measurement (Eve cannot tailor her attack to the measurement in each round individually, since she does not know beforehand which measurement will be chosen).

In the following, we first quantify the security of the protocol using the asymptotic secret key rate, $K_\infty$. This quantity is the ratio of the extractable secret key length to the total number of measurement rounds $N$, where $N \rightarrow \infty$. In the asymptotic limit, we may also take $q \rightarrow 1$, which maximises the so-called sifting factor~\cite{Lo2005} and get 
\begin{equation}\label{keyrates}
    K_\infty=p_{s} r_\infty,
\end{equation}
where $p_{s}:=p^2+(1-p)^2$ is the probability of having matching key bases, and $r_\infty$ is the secret fraction~\cite{scarani2009security}. The latter is given in terms of entropic quantities and reads

\begin{align}\label{rinfy}
    r_\infty:=&\underbrace{\lambda H(A_0\vert E)+\lt H(A_1\vert E)}_{H(Z\vert E\Theta)}\nonumber \\ &\quad-\lambda h(Q_{A_0B_0})-\lt h(Q_{A_1B_1}),
\end{align}
where $h(x):=-x\log(x)-(1-x)\log(1-x)$ is the binary entropy function,  $\lambda :=p^2/p_s$, $Q_{A_XB_Y}:=P(A_X\neq B_Y|X,Y)$ is the quantum bit error rate (QBER) for $X,Y$, and $E$ is Eve's quantum side-information gathered just before the error correction step. Here, $Z=A_\Theta$ and $\Theta \in\{0,1\}$ is the random variable denoting Alice's basis choice conditioned on the event either $X=Y=0$ or $X=Y=1$. Moreover, $E$ refers to quantum side information possessed by Eve. Hence, Eve's knowledge is fully described by $E \Theta$. The second line in equation~\eqref{rinfy} is the amount of information leaked to Eve during the error correction step (decoding with side-information $\Theta$).

The main challenge here is to put a reliable lower bound  on the conditional von Neumann entropy $H(Z\vert E\Theta)$, which measures the amount of uncertainty Eve has about $Z$ given side-information $E\Theta$, using solely the observed CHSH violation, $S$. To this end, we employ a family of device-independent entropic uncertainty relations~\cite{berta,eureview2}, which we can solve efficiently and reliably using a short sequence of numerical computations. More specifically, we seek to establish weighted entropic inequalities of the form
\begin{equation}
    \lambda H(A_0|E)+(1-\lambda)H(A_1|E)\geq C^*(S),
\end{equation}
where $C^*(S)$ is a function of the observed CHSH violation, $S$. A proof sketch is outlined in the Methods section and the complete analysis is provided in the accompanied Supplementary Note 1. 

A commonly used noise model for benchmarking the security performance of different QKD protocols is the depolarising channel model~\cite{pironio,acin2007device,scarani2009security}. In this noise model, all QBERs are the same and related to the CHSH value $S$ via
\begin{align}
    Q_{A_0B_0}=Q_{A_1B_1}=Q=\frac 12 \left(1 - \frac{S}{2\sqrt{2}}\right).\label{whiten}
\end{align}
Using this model, we compute the secret key rate and $H(Z\vert E\Theta)$, which are presented in Fig.~\ref{fig:ratesentro}. Here, $\lambda$ is a free parameter (i.e., a protocol parameter) that can be optimised by Alice and Bob (i.e., they optimise $p=P(X=0)$) for a given pair of $(S,Q)$. The result of this optimisation is remarkably simple: it is optimal to use a protocol with $\lambda =1/2$ (uniformly random key generation bases) if $S \lessapprox 2.5$( high noise) and  set $\lambda =1$, i.e., a fixed key generation basis otherwise (low noise). Surprisingly, there is no need to consider the intermediate values of $1/2<\lambda<1$.
In the case of the latter, our proposal reverts back to that of Ac\'{i}n et al.~\cite{mayers1998quantum,pironio,acin2007device} and the computed secret key rate appears to exactly match their analytical key rate bound. When using $H(Z|E\Theta)$ as a performance metric (which only depends on $S$ and thus applies to a general class of channel models satisfying this constraint), we observe that the uncertainty of Eve for our proposal is always higher than that of the original protocol for all $S\in(2,2\sqrt{2})$, see the right side of Fig.~\ref{fig:ratesentro}. In fact, for $\lambda=\frac{1}{2}$ our proposal is nearly optimal, in the sense that the bound on $H(Z|E\Theta)$ is very close to the linear bound, which is the fundamental upper limit of Eve's uncertainty given a fixed $S$. However, while it is optimal in this sense, choosing $\lambda=\frac{1}{2}$ is not always optimal in terms of producing the highest secret key rates, because the key rate is penalized by the sifting factor. Hence, $\lambda=1$ is preferred for the region $S \gtrapprox 2.5$.

\begin{figure*}[t]\centering
	\begin{minipage}[c]{0.5\linewidth}
	(a)\\
		\includegraphics[width=\linewidth]{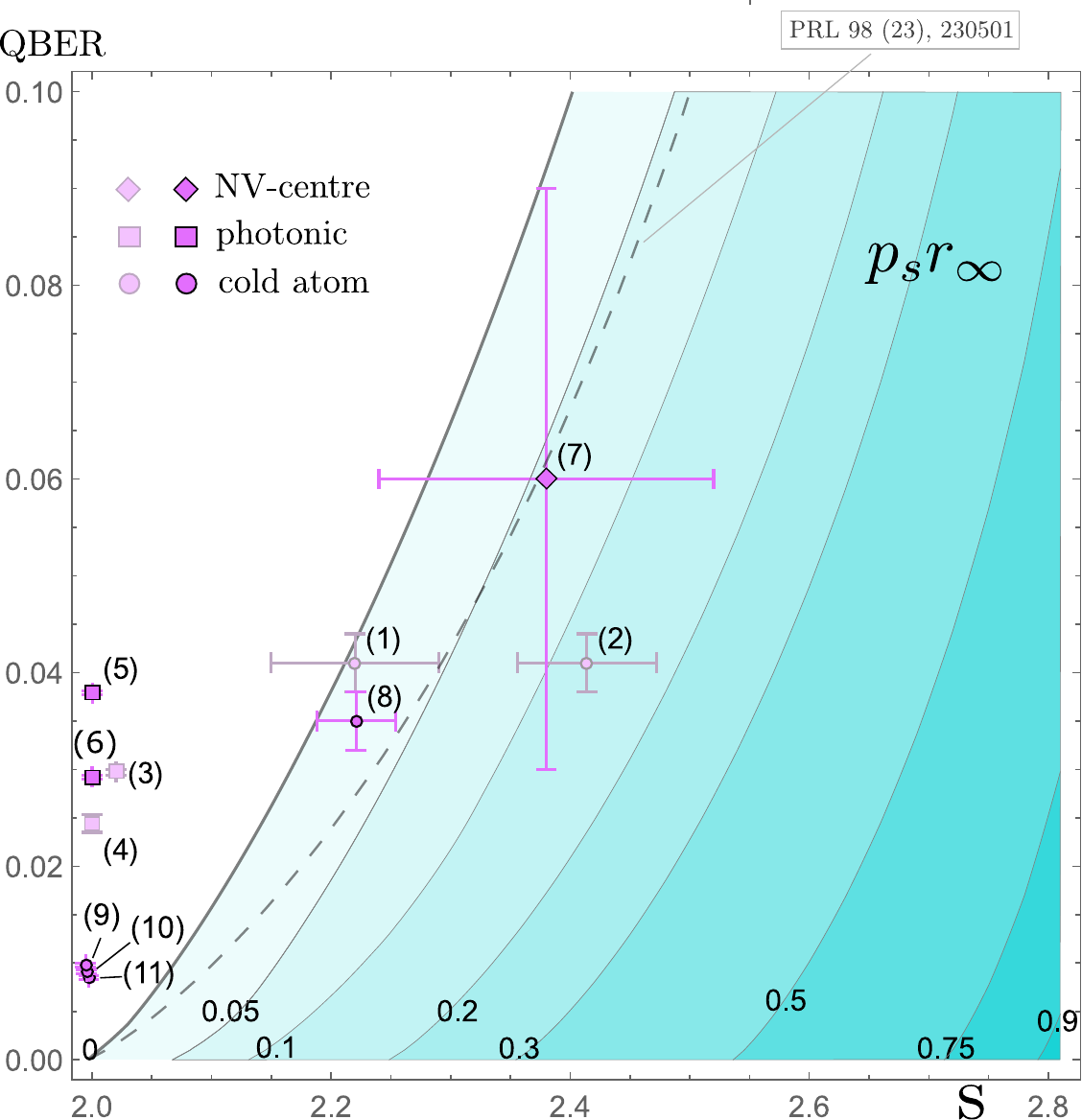}
	\end{minipage}
	\begin{minipage}[c]{0.45\linewidth}
	\hspace{-4.5cm}(b)\\
	  \includegraphics[width=0.9\textwidth]{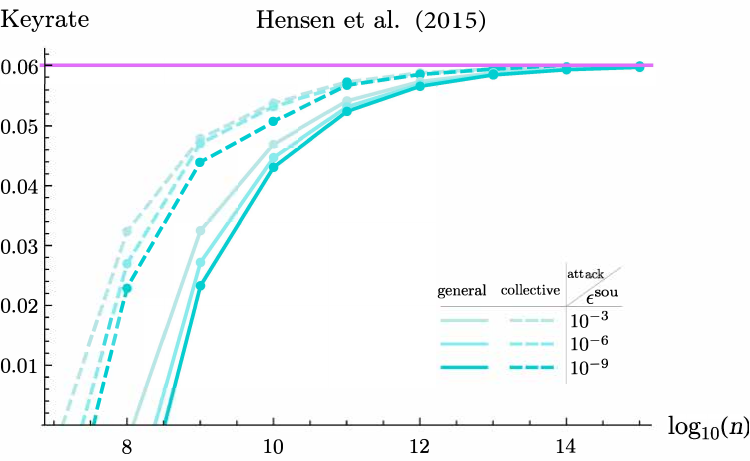}\\
	 \hspace{-4.5cm}(c)\\
      \includegraphics[width=0.9\textwidth]{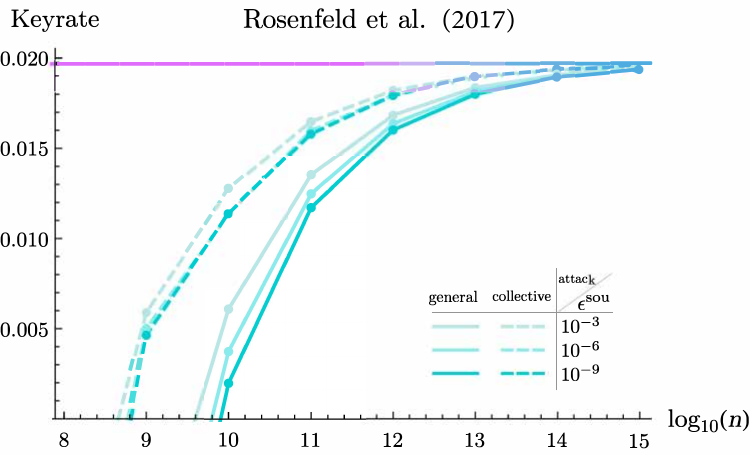}
	\end{minipage}\\%
	\newcommand{\eps}{\epsilon}
\newcommand{\ecom}{\eps^\mathrm{com}}
\newcommand{\esound}{\eps^\mathrm{sou}}

\caption{\label{fig:rates}\textbf{Rates for existing experiments:}  The contour plot (a) illustrates the asymptotic key rate $K_\infty=p_sr_\infty$ as a function of $S$ and $QBER$. We marked the location of recent experiments (see Tab.~\ref{tab:experiments}). Our DIQKD proposal suggests that now a positive asymptotic key rate of reasonable magnitude is possible for  experiments (7,8).  
The plots (b) and (c) show the finite-size key rates as a function of number of rounds, for the choice $p=1/2$ (which appears optimal at these noise levels).
These plots are for the estimated parameters in Murta et al.~\cite{Murta2018} for the Bell tests in (b) Hensen et al.~\cite{Hensen2015} and (c) Rosenfeld et al.~\cite{Rosenfeld2017} respectively.
The solid curves show the results for general attacks, while the dashed curves show the results under the assumption of collective attacks. The different colours correspond to different soundness parameters $\esound$ (informally, a measure of how insecure the key is; see Supplementary Note 4) as listed in the inset legends,
while the completeness parameter (the probability that the honest devices abort) is $\ecom=10^{-2}$ in all cases. The horizontal line denotes the asymptotic key rate. Note that only experiments (5,6,7,8,9,10,11) are loophole-free Bell tests, closing both detection and locality loopholes. On the other hand, experiments (1,2,3,4) did not close the locality loophole.
}
\end{figure*}

\begin{table}[h!]
\caption{\label{tab:experiments} \textbf{Asymptotic rates for existing experiments:} (data from Murta et al.~\cite{Murta2018}, Tab.~4).  Non-photonic experiments ((1,2) and (7,8)) now promise a positive keyrate. However, note that the experiments (1,2) were performed in a single lab and therefore did not close the locality loophole. The photonic experiments (3,4) also did not close the locality loophole.  The more recent experiments (5,6,7,8,9,10,11) closed both locality and detection loopholes. 
	    Note that the value of QBER for experiment (8) provided by Murta et al.~\cite{Murta2018} is based on the experiment of Henkel et al.~\cite{henkel2010highly} while the QBER achievable by experiment (8) is estimated to be higher by the authors of experiment (8)~\cite{Rosenfeld2017}.}
	    \centering
	  \begin{tabular}{cc|lccc|c}
	    Label && Experiment && Year && \qquad Rate [bit]\\\hline
	        (1) && Matsukevich et al.
		~\cite{matsukevich2008bell}& Cold atom &2008 && $0.004 $\\
	    	(2) && Pironio et al.,
		~\cite{pironio2010random}  &Cold atom &2010&& $0.118 $  \\
		
	        (3) && Giustina et al.	~\cite{giustina2013bell}&Photonic & 2013&&  $0$\\
	        (4) &&Christensen et al.
		~\cite{christensen2013detection}  &Photonic& 2013&& $0$\\
	        (5) && Giustina et al.
		~\cite{Giustina2015} &Photonic& 2015 && $0$\\
	        (6)&& Shalm et al.
		~\cite{Shalm2015} &Photonic & 2015&&  $0$\\
	        (7) && Hensen et al.
		~\cite{Hensen2015} &NV-centre& 2015&& $0.057$\\
	        (8) && Rosenfeld et al.
		~\cite{Rosenfeld2017} &Cold atom& 2017&& $0.019$\\
			(9) && Liu et al.
		~\cite{Liu18} &Photonic& 2018&& $0$\\
			(10) && Liu et al.
		~\cite{Liu19} &Photonic & 2019&& $0$\\
			(11) && Li et al.
		~\cite{Li19} &Photonic& 2019&& $0$
	    \end{tabular}

\end{table}

In order to evaluate the feasibility of our proposal, we look at the existing list of loophole-free CHSH experiments~\cite{Murta2018} and compute the corresponding secret key rates. Generally speaking, there are two types of Bell experiments: one based on measuring entangled photon pairs using high efficiency single-photon detectors, and the other based on event-ready systems~\cite{zukowski1993event} using entanglement swapping between entangled photon pairs and atoms/NV-centres. Following along the lines of Murta et al.~\cite{Murta2018}, we prepare a feasibility region plot for the list of CHSH experiment therein, which is presented in Fig.~\ref{fig:rates}. The immediate observation is that our DIQKD proposal significantly expands the region of channel parameters that give rise to positive key rates, thus substantially improving the robustness of DIQKD. The next observation is that event-ready loophole-free CHSH experiments~\cite{Hensen2015,Rosenfeld2017} are now well within the positive key region; as opposed to the original protocol where they are either in the insecure region or around the boundary. Unfortunately, CHSH experiments based on entangled photon pairs are still in the insecure region (also see Supplementary Note 2), although it should be mentioned that this observation holds only for our proposal and the original DIQKD protocol.

Our results hence show that positive asymptotic key rates can be achieved by recent event-ready loophole-free experiments.
This significantly improves over the original protocol~\cite{pironio}, which does not achieve positive key rates for any such experiments, even in the asymptotic limit (though Murta et al.~\cite{Murta2018} describes prospective future improvements to NV-centre implementations, which may allow positive asymptotic rates). 
However, there are still a few experimental challenges. For one, we note that the event-ready CHSH experiments are fairly slow compared to their photonic counterparts; e.g.~the event-ready experiment by Hensen et al.~\cite{Hensen2015} performed only 245 rounds of measurement during a total collection time of 220 h. 
Recently, Humphreys et al.~\cite{humphreys2018deterministic} demonstrated that it is possible to improve the entanglement rate by a couple of orders of magnitude, but this comes at the expense of the overall state fidelity and hence lower CHSH violations. 
While our protocol can yield positive asymptotic key rates in these noise regimes, a relevant question to consider is the number of rounds required to achieve security in a finite-key analysis.

To make this concrete, we analyse the finite-key security of our protocol using the proof technique from a recent work~\cite{TSB+20} 
(see the Supplementary Note 4 for the proof sketch). In particular, we compute the finite-key rates for both collective and general attacks, with the analysis of the latter making use of the entropy accumulation theorem~\cite{eat,eatqkd,eatsecondorder}, which essentially certifies the same asymptotic rates as in the collective attacks scenario. (An alternative approach may be the quantum probability estimation technique~\cite{Zhang2020Efficient}.) Our results are summarised in Fig.~\ref{fig:rates}, focusing on the experiments from Hensen et al.~\cite{Hensen2015} and Rosenfeld et al.~\cite{Rosenfeld2017} (which can achieve positive asymptotic key rate, as mentioned above). We see from the plots that they require approximately $10^8$ and $10^{10}$ measurement rounds respectively to achieve positive finite-size key rates against general attacks. In these experimental implementations, this number of rounds is still currently out of reach (assuming realistic measurement time). Overall, however, given future improvements on these experimental parameters, our protocol would attain higher asymptotic rates than the original protocol~\cite{pironio}, and hence also require fewer rounds to achieve positive finite-key rates.

To further improve the robustness and key rates, there are a few possible directions to take. For one, we can consider the full input-output probability distribution instead of just taking the CHSH violation. Since the latter only uses part of the available information, more secrecy could potentially be certified by finding methods to compute secret key rates that take into account the full probability distribution estimated from the experiment. 
Such a method for general Bell scenarios was recently developed~\cite{ourworkarxiv}; however, we found (see Supplementary Note 1) that the bounds it gives in this case are not tight, and are slightly worse than the results presented above. Another possible approach specialized for 2-input 2-output scenarios is presented in Tan et al.~\cite{TSB+20}, which is potentially more promising for such scenarios.\newline
 
\section*{Methods}
\label{methods}
\noindent Here, we outline the main ideas of our security analysis. The core of the security analysis is a reliable lower-bound estimate on the conditional von Neumann entropy of Eve. The complete analysis is deferred to the accompanying Supplementary Note 1.

\subsection*{Average secret key rate}\label{keyratecomp}
\noindent Conditioned on the key generation rounds, Alice and Bob would pick their inputs (basis choices) according to a probability distribution $(p,1-p)$. As discussed in the main text, this distribution acts as a free parameter in our protocol and has to be adapted to a given set of channel parameters $(S,Q)$ in order to obtain an optimal performance.  
In the following we will therefore outline how the final key rate $K_\infty$ and secret fraction $r_\infty$ are given as functions of $(p,S,Q)$. 

The secret fraction in a round of the protocol, in which the measurements $A_X$ and $B_Y$ are obtained, can be computed using the Devetak-Winter bound ~\cite{devetak} under the assumption of collective attacks,
\begin{equation}
    r_\infty^{A_X B_Y}
    \geq H(A_X|E)-H(A_X|B_Y).
    \label{dwbound}
\end{equation}
Here the term $H(A_X|B_Y)$ only depends on the statistics of the measured data of Alice and Bob, and can therefore be directly estimated in an experiment. For binary measurements this quantity can be furthermore expressed  by the respective bit error rate $Q_{A_X B_Y}$ via
\begin{equation}
    H(A_X|B_Y)\leq h(Q_{A_XB_Y}).
\end{equation}
In our protocol, a key generation round is obtained whenever Alice and Bob perform measurements $A_X$ and $B_{Y}$ with $X=Y$. The probability that Alice and Bob perform the measurements $A_0$ and $B_0$ is $p^2$ and the probability that they perform $A_1$ and $B_1$ is $(1-p)^2$.  When the error correction is done for both cases, ($A_0,B_0$) or ($A_1,B_1$), separately, we obtain the overall asymptotic key rate as sum of the individual secret fractions weighted by their respective probability. This gives 
\begin{align}
    K_\infty
    & \geq p^2 r_\infty^{A_0B_0}+(1-p)^2 r_\infty^{A_1B_1}\nnnl
    &=p^2 H(A_0\vert E) + (1-p)^2 H(A_1\vert E)\nnnl
    &\quad- p^2 H(A_0\vert B_0) - (1-p)^2H(A_1\vert B_1)\nnnl
    & \geq p_{s}\bigl(\lambda H(A_0\vert E)+\lt H(A_1\vert E)\nnnl
    &\quad \hspace{0.5cm}-\lambda h(Q_{A_0B_0})-\lt h(Q_{A_1B_1})\bigr)\nnnl
    &:=p_{s}r_\infty,
    \label{keyratesm}
\end{align}
where the success probability $p_s$ and the relative distribution of the basis choices $(\lambda,\lt)$ are given by
\begin{align}
    p_{s}
    &=\left(p^2+(1-p)^2\right)= 1-2p+2p^2
\end{align}
and
\begin{align}
    \lambda=\frac{p^2}{1-2p+2p^2},
\end{align} respectively. As mentioned in the main text we also write
\begin{align}
   H(Z\vert E\Theta):=\lambda H(A_0\vert E)+\lt H(A_1\vert E)\label{entrosubst}
\end{align}
where $\Theta$ denotes a binary random variable (distributed by $(\lambda,\lt)$) that (virtually) determines which basis pair, $(A_0,B_0)$ or $(A_1,B_1)$, is picked in a successful key generation round in order to generate the values of a combined random variable $Z=A_\Theta$. 

\subsection*{Device-independent entropic uncertainty relation}\label{eucrsec}

\noindent The only term in the key rate formula \eqref{keyratesm} that cannot be directly obtained from the measurement data is the conditional entropy $H(Z\vert E\Theta)$. The main challenge here is thus to establish a reliable lower bound on this quantity assuming only the CHSH violation $S$. More specifically, we are interested in finding a function $C^*(S)$ such that  
\begin{align}
 \quad H(Z\vert E\Theta)\geq C^*(S)\label{eucreq}
\end{align}
holds for \textit{all} possible combinations of states and measurements (in any dimension) that are consistent with the observed CHSH value $S$. An inequality like equation \eqref{eucreq} is commonly referred to as an entropic uncertainty relation, and in our case we are interested in relations with quantum side-information~\cite{berta,liebfrank,junge}. There is a vast amount of literature~\cite{eureview1,eureview2,thesis} in which relations of this form~\cite{berta,liebfrank,junge,liu} or similar~\cite{deutsch,hans2,muff,ourentro,schneeloch,additivity,alberto1} have been studied and several types of uncertainty relations have been discovered. A typical family of entropic uncertainty relations, which is close to our problem, is that proposed by Berta et al.~\cite{berta} and the weighted generalisation of it from Gao et al.~\cite{junge}. These inequalities, however, are not device-independent and require the measurement characterisation of at least one party, which unfortunately is not possible in our setting.

To the best of our knowledge, the only known entropic uncertainty relations for uncharacterised measurements are given by Tomamichel et al.~\cite{Tomamichel2013} and Lim et al.~\cite{limdevice}. There, the uncertainty of the measurement outcomes with side-information is lower bounded by the so-called overlap of the measurements~\cite{berta}, which in turn is further bounded by a function of the CHSH violation. Although these relations are applicable to uncharacterised measurements, they appear fairly weak when applied to our DIQKD proposal, i.e., they do not provide any improvement in the secret key rate when compared to the original protocol.

\begin{figure*}[t]
   \hspace{-2cm}(a) \hspace{8cm} (b)\\%
    \begin{minipage}[t]{0.36\linewidth}
     \vspace{0pt}
    \includegraphics[width=1\linewidth]{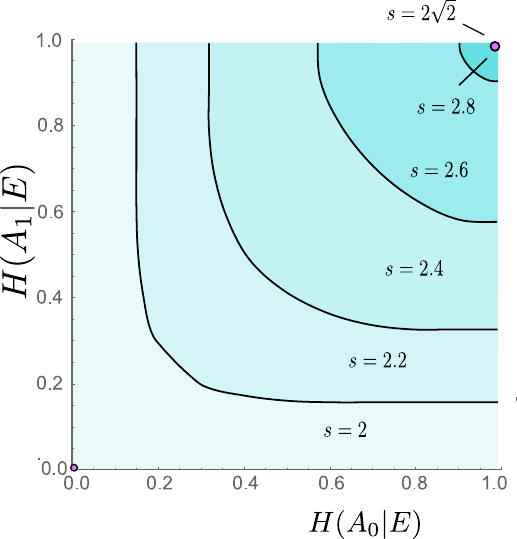}
    \end{minipage}\hspace{1cm}
    \begin{minipage}[t]{0.53\linewidth}
    \vspace{0pt}
    \includegraphics[width=1\linewidth]{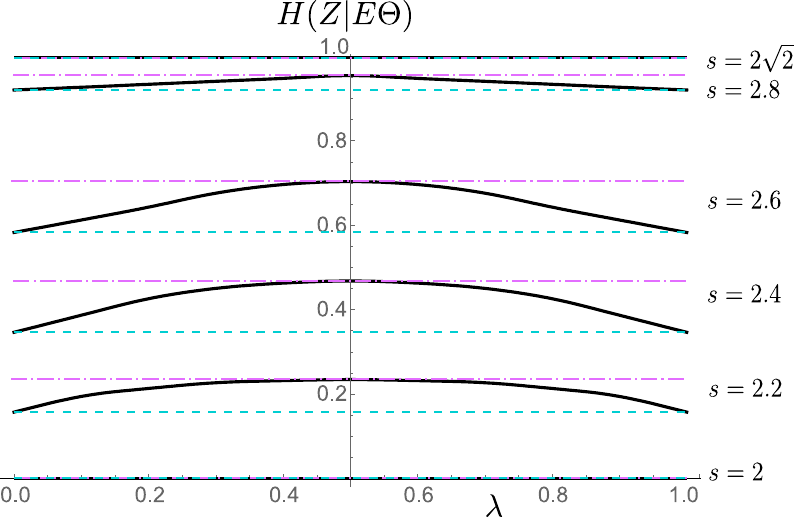}
    \end{minipage}
    \caption{\textbf{Device-independent uncertainty relations:}
    \label{fig:weights} In panel (a), the plot shows the device-independent uncertainty relation between $\lambda H(A_0|E)+(1-\lambda)H(A_1|E)$, where the solid line is the fundamental uncertainty $C^*(S)$ for a given CHSH violation. The shaded region above the line hence represents the feasible region of $(H(A_0|E),H(A_1|E))$ given $S$. Evidently, when $S=2\sqrt{2}$, we see that $H(A_0|E)=H(A_1|E)=1$ must be maximally random; indeed, $S=2\sqrt{2}$ corresponds to the case where Eve is completely uncorrelated with the devices and hence her best guess is limited to a random guess. In panel (b), the plot shows the minimal uncertainty $C^*(S)$ (the bottom dashed line) attained when $\lambda=0,1$ and one can see that $C^*(S)$ (solid line) for $0<\lambda <1$ always gives a non-trivial advantage over the limiting case (like in the original DIQKD protocol).    }
\end{figure*}

The lower bound we establish in this work, i.e., the function $C^*(S)$, appears to be optimal in that it can be saturated by two-qubit states up to numerical precision (see Supplementary Note 1). $C^*(S)$ is depicted in Fig.~\ref{fig:ratesentro}(b) for $\lambda=0.5$, and in Fig.~\ref{fig:weights}(b) for continuous values of $\lambda$ and $S\in\{2,2.2,2.4,2.6,2.8,2\sqrt{2}\}$. In Fig.~\ref{fig:weights}(a), we additionally plot the so-called uncertainty sets~\cite{amu,thesis,additivity} of our relation. These are sets that outline all the admissible pairs of entropies $(H(A_0\vert E),H(A_1\vert E))$ for a given lower bound on $S$. We also note that it may seem plausible to independently optimise each term $H(A_0\vert E)$, $H(A_1\vert E)$, instead of the sum of them. However, as shown in Fig.~\ref{fig:weights}(b), numerical results suggest that optimising the weighted sum of these terms is always better than optimising the individual terms: this also highlights where our DIQKD proposal improves over the original protocol. 

\subsection*{Computing \texorpdfstring{$C^*(S)$}{lower bound}}
\noindent As mentioned, the complete analysis of our lower bound on $C^*(S)$ is deferred to the Supplementary Note 1. In the following we will only outline the main steps of the analysis to reduce the computation of $C^*(S)$ to a sequence of problems that can be treated successively. The basic idea to find a way to compute $C^*(S)$ by only using known estimates (both analytical and numerical) that give a final lower bound that is reliable. This means that all the steps of the analysis assume the worst-case scenario, so that final value is a strict lower bound on $C^*(S)$. In brief, the analysis uses a refined version of Pinsker's inequality, semi-definite optimisation and an $\e$-net to achieve this goal. 

Our first step is to reformulate the tripartite problem involving Alice, Bob and Eve to a bipartite one that involves only Alice and Bob. Following Tan et al.~\cite{ourworkarxiv}, we note that conditional entropy terms like $H(A_0|E)$ can always be reformulated as the entropy production $H(T_X(\rho_{AB}))-H(\rho_{AB})$ of the quantum channel $T_X$ on the post measurement state on the Alice-Bob system, which is defined as the change of von Neumann entropy of the system that is subjected to the quantum channel $T_X$. In our case, this channel $T_X$ is a pinching channel, defined as:
\begin{equation}
    T_X[\rho]:=(\Pi^{A_x}_0\otimes\id) \,\rho\, (\Pi^{A_x}_0\otimes\id) + (\Pi^{A_x}_1\otimes\id) \,\rho\, (\Pi^{A_x}_1\otimes\id),
\end{equation}
where $\Pi^{A_x}_a$ denotes the projector associated with Alice's measurement setting $x$ and outcome $a$. Clearly, the pinching channel satisfies $T_X=T_X^2=T_X^*$ and acts complementary to the map that models Alice's measurement. With this, we can further rewrite the entropy production as
\begin{align}
&\lambda H(A_0|E)+\lt H(A_1|E)\nnnl
&=\lambda H(T_{0}[\rho_{AB}])-\lambda H(\rho_{AB}) \nnnl%
&\quad+ \lt H(T_{1}[\rho_{AB}])- \lt H(\rho_{AB})\nnnl
&= \lambda D(\rho_{AB}\Vert T_0[\rho_{AB}])+ \lt D(\rho_{AB}\Vert T_1[\rho_{AB}])\label{condiboundM},
\end{align} where $D(\rho \Vert \sigma)$ is the quantum relative entropy of $\rho$ with respect to $\sigma$.

\begin{figure}
    \centering
    \includegraphics[width=0.92\linewidth]{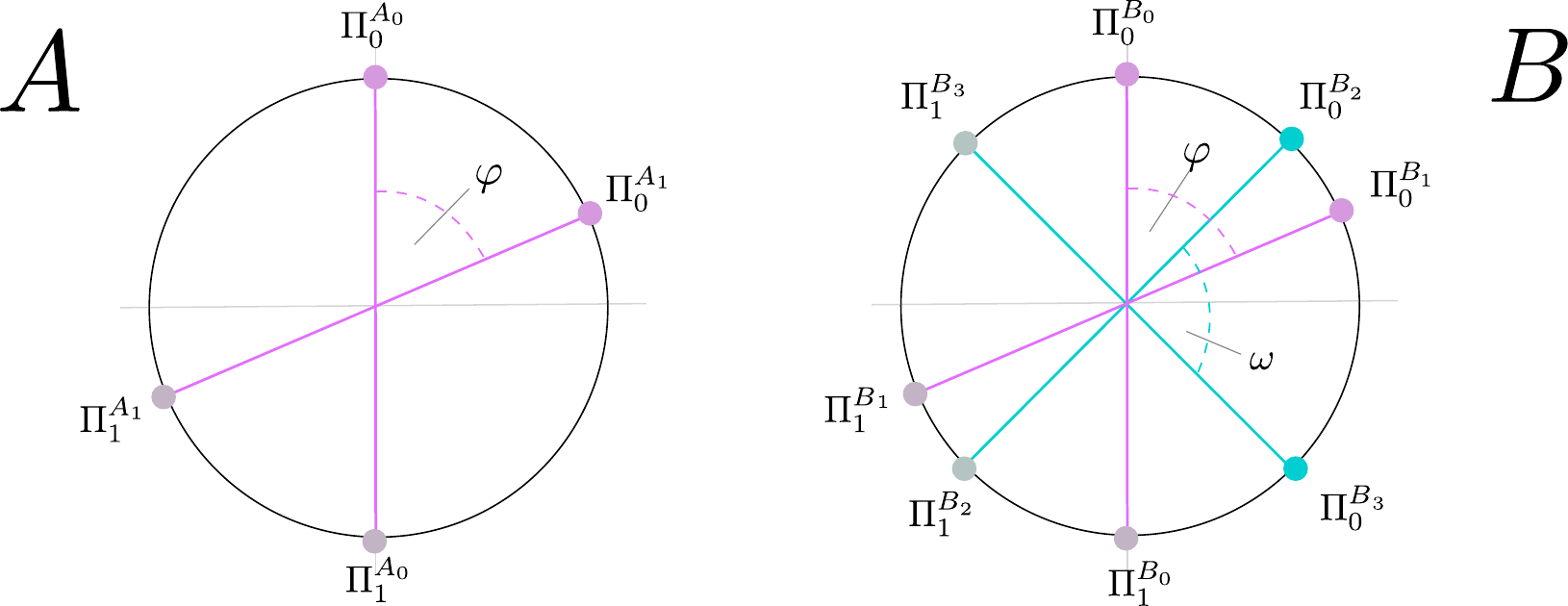}
    \caption{\textbf{Measurement setting for two-qubit states:}
    Alice has two projective measurement that are described by projectors $(\Pi_0^{A_0},\Pi_1^{A_0})$ and 
    $(\Pi_0^{A_1},\Pi_1^{A_1})$ with relative angle $\varphi$. Bob has in total four measurements: for key generation, ideally he should perform measurements $B_0$ and $B_1$ which are aligned with Alice's measurements $A_0$ and $A_1$ to minimise the quantum bit error rates. The security analysis only depends on Bob's measurements $B_2$ and $B_3$, which are w.l.o.g. specified by a relative angle $\omega$. }
    \label{fig:measurement}
\end{figure}

Then, we follow a proof technique in the original work on DIQKD~\cite{pironio,acin2007device} to reduce the underlying $\rho_{AB}$ to a mixture of two-qubit states, where it is assumed that the mixing is due to Eve. That is, since each party (Alice and Bob) performs only two binary measurements, their local measurement devices can be described by only specifying two projectors (whose dimensions are unspecified). The corresponding local algebras, which are generated by two projectors, are  well investigated mathematical objects~\cite{bottcher} for which a central theorem~\cite{halmos} states that their representation can be decomposed into $2\times2$ (qubit) blocks and a commuting rest. Correspondingly, this allows us to conclude (details in the Supplementary Note 1) that the desired uncertainty bound can be decomposed accordingly as a convex combination:
\begin{align}\label{convextqM}
C^*(S)
    &\geq\inf_{\mu}\quad \int_{S'=2}^{2\sqrt{2}}\mu(dS')\, C^*_{\mathbb{C}^{4\times4}}(S') \\
    &\quad \textnormal{s.t.}\quad  \mu([2,2\sqrt{2}]) \leq 1 ,\quad \mu\geq0 \nnnl
    &\quad\quad\quad  \quad  \int_{S'=2}^{2\sqrt{2}}\mu(dS')S'=S,\nonumber
\end{align}
where $C^*_{\mathbb{C}^{4\times4}}(S')$ is a lower bound on the conditional entropy $H(Z|E\Theta)$ for projective measurements on two qubits. Here, we note that once a bound on $C^*_{\mathbb{C}^{4\times4}}(S')$ is established, the optimisation over all measures $\mu$, which can be geometrically interpreted as taking a convex hull, is straightforward to perform. 
As shown in Fig.~\ref{fig:measurement}, the situation for the optimisation corresponding to $C^*_{\mathbb{C}^{4\times4}}(S')$ can now, w.l.o.g., be fully described by specifying a two-qubit state and two angles $(\varphi,\omega)$ that describe the relative alignment of Alice's and Bob's measurements. 

Although the problem has been reduced to two-qubit states and projective qubit measurements, a direct computation of $C^*_{\mathbb{C}^{4\times4}}(S')$ is still an open problem as there are no known proof techniques that can be applied to our situation. To that end, we employ a refined version of Pinsker's inequality (see Theorem 1 in the Supplementary Note 3), 
\begin{align}\label{PinskerM}
D(\rho\Vert T[\rho])\geq \log(2)-h_2\left(\frac 12 - \frac 12 \left\Vert \rho-T[\rho]\right\Vert_1 \right),
\end{align}
to obtain a lower bound on the relative entropy in \eqref{condiboundM} in terms of the trace norm.

The big advantage of establishing estimates in terms of the trace norm is that a minimisation thereof can be formulated as a semi-definite program (SDP). (In fact, it is possible in principle to use this inequality to bound the entropy without reducing the analysis to qubits, though the resulting bounds in that case do not appear to be very tight. We discuss this in detail in Supplementary Note 1.)

With that, the overall optimisation problem at hand now reads
\begin{align}
\label{linopti2M}
\inf_{\varphi \in[0,\pi/2]} ~
\inf_{\mathbf{b}: \Vert\mathbf{b}\Vert_2=1 }\quad \left\{
\begin{matrix}
\inf_{\rho} & \lambda  \delta\left(\rho,0\right) + \lt  \delta\left(\rho,\p\right)\\ &\\
 \textnormal{s.t.} & \quad \erw{F_0+\mathbf{F}\cdot \mathbf{b} }_{\rho}=S
\end{matrix} \right\},
\end{align}
where
\begin{align}
    \delta\left(\rho,\p\right):=\Bigl\Vert \bigl\{\rho,Q(\p)\bigr\}-2 Q(\p)\rho Q(\p)\Bigr\Vert_1,
\end{align}
and constraints that are linear in $\rho$ given by $4\times 4$-matrices  
\begin{align}
    F_0 \und \mathbf{F}\cdot \mathbf{b}=b_xF_x(\p)+b_zF_z(\p),
\end{align}
where $Q(\p),F_x(\p)$ and $F_y(\p)$ depend on $\p$ in terms of the first and second order in $\cos(\p)$ and $\sin(\p)$. 
In the above expression, $\mathbf{b}$ is a vector on a (2-norm) unit sphere that arises from reformulating the description of Bob's measurements. 

\def\gbrk #1{{\color{grun}(\color{Black}#1\color{grun})}}%
\def\mbrk #1{{\color{magenta}(\color{Black}#1\color{magenta})}}%

This optimisation can be solved in three stages as indicated in \eqref{linopti2M}:
\begin{itemize}
    \item[(i)] The rightmost optimisation is an SDP on $4\times 4$ matrices, which can be efficiently solved~\cite{boyd04p8}. 
    \item[\mbrk{ii}] The second optimisation is performed by relaxing the continuous optimisation over the (2-norm) unit sphere to a discrete optimisation on a sequence of polygonial approximation (similar to the method used in Schwonnek et al.~\cite{varucr1,varucr2}). Also this optimisation can be performed with reliable lower bounds to the order of any target precision.
    \item[\gbrk{iii}] The leftmost optimisation only runs over the single parameter $\p$ coming from a bounded domain. Hence, it is possible to efficiently tackle this optimisation by an $\e$-net. In order to do so it is required to provide an error estimate (for the magenta and the black box) for all $\p$ that are located in an $\e$-interval around some $\p_0$. Note that all previous optimisations are linear in $\rho$ and $\mathbf{b}$ and only depend in the second order on $\cos(\p)$ and $\sin(\p)$ (which are bounded functions of $\p$). 
\end{itemize}

\noindent{\textbf{Acknowledgements}}\newline
R.~S., K.~T.~G., and C.~C.-W.~L. were funded by the National Research Foundation of Singapore, under its NRF Fellowship grant (NRFF11-2019-0001) and NRF Quantum Engineering Programme grant (QEP-P2), and the Centre for Quantum Technologies. V.~S. and I.~W.~P. were supported by NRF and the Ministry of Education, Singapore, under the Research Centres of Excellence program.
E.~Y.-Z.~T.~was funded by the Swiss National Science Foundation via the National Center for Competence in Research for Quantum Science and Technology (QSIT), the Air Force Office of Scientific Research (AFOSR) via grant FA9550-19-1-0202, and the QuantERA project eDICT. R.~W. was supported, in part, by the DFG through SFB 1227 (DQmat), the RTG 1991, and funded by the Deutsche Forschungsgemeinschaft
(DFG, German Research Foundation) under Germany’s Excellence Strategy EXC-2123 QuantumFrontiers 390837967.

\newpage
\begin{widetext}


\section*{Supplementary material}

# Supplementary information: Device-independent quantum key distribution with random key basis

René Schwonnek,[1, ∗] Koon Tong Goh,[1, ∗] Ignatius W. Primaatmaja,[2] Ernest Y.-Z. Tan,[3] Ramona Wolf,[4] Valerio Scarani,[2, 5] and Charles C.-W. Lim[1, 2]

[1] *Department of Electrical & Computer Engineering, National University of Singapore, Singapore*
[2] *Centre for Quantum Technologies, National University of Singapore, Singapore*
[3] *Institute for Theoretical Physics, ETH Zürich, Switzerland*
[4] *Institut für Theoretische Physik, Leibniz Universität Hannover, Germany*
[5] *Department of Physics, National University of Singapore, Singapore*

## I. SUPPLEMENTARY NOTE 1 - SECURITY ANALYSIS

This section documents the derivation of our lower bound on the quantity $C^*(S)$, which is introduced in the main text. $C^*(S)$ is defined as the minimum $H(Z|E\Theta)$ (conditional entropy of $Z$ given side-information $E\Theta$) upon observing a CHSH value of $S$. Recall that

$$H(Z|E\Theta) := \lambda H(A_0|E) + (1-\lambda) H(A_1|E) \tag{1}$$

describes Eve's uncertainty about the measurement outcome $Z$ given $\Theta$ and is a keystone for determining the secret key rate of our device-independent quantum key distribution (DIQKD) protocols.

Before diving into the derivation, we will furnish the definitions and notations of the quantum states, operators and functionals that will be used throughout this section. In the main text, we considered a Bell scenario where two parties Alice and Bob input measurement settings $X \in \{0,1\}$ and $Y \in \{0,1,2,3\}$ respectively into their devices and in return, each device outputs a bit. We shall omit the measurement settings $Y = 0, 1$ for the entirety of this section as they do not serve any role in the estimation of $H(Z|E\Theta)$.

Since we are working in a fully device-independent setting, we can assume that all measurements can be written as projectors without any loss of generality (refer to the explanation in [1]). As such, we denote the projectors corresponding to measurements $A_0, A_1, B_2, B_3$ as

$$A_0 = \{\Pi_0^{A_0}, \Pi_1^{A_0}\}, \quad A_1 = \{\Pi_0^{A_1}, \Pi_1^{A_1}\}, \quad B_2 = \{\Pi_0^{B_2}, \Pi_1^{B_2}\}, \text{ and } B_3 = \{\Pi_0^{B_3}, \Pi_1^{B_3}\}, \tag{2}$$

where the subscript of each projector denotes its corresponding measurement outcome. Therefore, the correlation measurement operators are defined by

$$C^{A_0} := \Pi_0^{A_0} - \Pi_1^{A_0}, \quad C^{A_1} := \Pi_0^{A_1} - \Pi_1^{A_1}, \quad C^{B_2} := \Pi_0^{B_2} - \Pi_1^{B_2}, \text{ and } C^{B_3} := \Pi_0^{B_3} - \Pi_1^{B_3}. \tag{3}$$

The correlation measurement operators and the correlation function $C_{XY} = P(A_X = B_Y|X,Y) - P(A_X \neq B_Y|X,Y)$ defined in the main text is related via $C_{XY} = \text{tr}\left(\rho\, C^{A_X} \otimes C^{B_Y}\right)$ where $\rho$ is the measured quantum state. Hence, the CHSH operator is denoted as

$$CHSH := C^{A_1} \otimes C^{B_2} - C^{A_0} \otimes C^{B_2} - C^{A_0} \otimes C^{B_3} - C^{A_1} \otimes C^{B_3}. \tag{4}$$

Furthermore, we denote the single round quantum state of Alice and Bob by $\rho_{AB}$ and assume that Eve holds the purification of this state. Thus, the joint state describing Alice, Bob, and Eve quantum systems is given by some pure state $\psi_{ABE}$, whose Hilbert space dimension is unknown. Conventionally, the von Neumann entropy of a quantum state $\rho$ is denoted and defined by $H(\rho) := -\text{tr}(\rho \log \rho)$, we shall adopt the shorthand notation of

$$H(AB) = H(\rho_{AB}), \quad H(B) = H(\rho_B), \quad H(E) = H(\rho_E) \ldots \tag{5}$$

to denote the entropy of a specific subsystem, whenever the underlying state is clear from the context. Next, a classical-quantum state after the action of Alice's measurement, $A_X$, is modelled by

$$\rho_{A_X BE} = \sum_{i \in \{0,1\}} |i\rangle\langle i| \otimes \text{tr}_A\left((\Pi_i^{A_X} \otimes \mathbf{1}_{BE})\psi_{ABE}(\Pi_i^{A_X} \otimes \mathbf{1}_{BE})\right), \tag{6}$$

---

[∗] These authors contributed equally to this work.

where $|i\rangle\langle i|$ indexes a classical register that stores the measurement outcomes. Thus, the conditional entropy $H(A_X|E) = H(A_X E) - H(E)$ is well-defined for any given joint quantum state describing Alice, Bob and Eve's systems through the above-mentioned relations.

Finally, we could state the proper mathematical definition of $C^*(S)$ using the furnished definitions:

$$C^*(S) = \inf_{A_0,A_1,B_2,B_3} \inf_{\psi_{ABE}} \lambda H(A_0|E) + (1-\lambda)H(A_1|E)$$
$$\text{s.th.:} \quad \langle CHSH\rangle_{\rho_{AB}} = S. \tag{7}$$

In the following subsections, we will provide the detailed derivation of our lower bound on the quantity $C^*(S)$ and the structure of the proceedings are organised as follows:

I A We start by following a well-known argument [1–3], which allows us to reformulate the estimation of conditional entropy of Eve in a tripartite scenario into the estimation of **entropy production on the Alice-Bob system**.

I B We proceed by following the argument of Pironio et al.[4], which allows us to conclude that the worst case, i.e. the optimal attack of Eve, can always be realised by Alice and Bob performing a convex combination of **projective measurements on a two qubit state**. Even though the problem now may seem straightforward, it still requires the application of a suite of non-trivial numerical and analytical methods. This is due to the fact that the simultaneous optimisation over states and measurements is neither linear nor convex.

I C We then apply a new refined version of Pinsker's inequality (see Thm. 1) that enables the estimation of entropies via certain trace norms. The remaining problem of **optimising over trace norms** can then be bounded using an efficient algorithm, which will be introduced in the follow subsections.

I D We first **formulate the optimisation over the unknown quantum state as a semi-definite program (SDP)**.

I E Although the **optimisation over Bob's measurements** is not convex, we formulated an efficient algorithm that lower bounds this optimisation up to arbitrary precision.

I F At this stage, it only remains to **optimise over Alice's measurements**. Similar to the previous subsection, optimising over Alice's measurements also result in an optimisation problem that is not a SDP. Since this optimisation only involves a single parameter, which is the angle $\varphi$ between Alice's measurements, we could simply employ an adaptive refining $\varepsilon$-net. This is possible as all operators involved in the previous steps can be explicitly bounded in norm.

I G At this point we are able to compute reliable lower bounds on $C^*(S)$ for any two-qubit strategy. Using I B, we are able to compute the final $C^*(S)$ for a discretised range of different Bell violations $S$ on qubits and taking its convex hull subsequently.

### A. Reduction to locally accessible quantities

The defining advantage of quantum over classical cryptography stems from the fact that for a quantum system it is possible to witness the action of Eve, given only access to the Alice-Bob subsystem. In this subsection, we will relate the conditional entropy $H(A_X|E)$, which depends on Eve's system, to quantities that only depends on Alice and Bob's systems. Such effort will allow us to circumvent the obstacle of having to perform optimisation over Eve's quantum system with unbounded Hilbert space dimension.

We begin by viewing any key generating measurement, $A_X$, of the DIQKD protocol from the perspective of a larger Hilbert space (similarly to the appendix of [2]) by introducing a classical memory $\mathcal{A}_X$ that is used to store the measurement outcome. We denote $\mathcal{A}_X^{init}$ as the abelian algebra describing the initial (empty) state of the memory and $\phi$ as its associated pure quantum state. As such, the process of measuring $A_X$ on Alice's quantum system (modelled by von Neumann algebra $\mathcal{A}$) and recording the outcome into the memory can be described by a unitary evolution that transforms the pure initial state $\Psi_{ABE} \otimes \phi$ into a final state $\Psi_{A'BEA_X}$. Note that the unitary evolution also involves the transformation of $\mathcal{A}$ into the post measurement state of a system $\mathcal{A}'$. Hence, $H(A_X|E)$ can be understood as entropy production $\Delta H$ on the memory-Eve system.

From the dilated perspective (see Fig. 1), the memory-Eve system is only one side of a bipartition, where the Alice-Bob system is the other. Since the local von Neumann entropies on two sides of a bipartite pure state are equal (see Thm.2(c) of [5]), the entropy production on the Alice-Bob system is equal to the entropy production of the memory-Eve system. Consequently, we can conclude that

$$H(A_X|E) = H(A_X E) - H(E) = H(A_X E) - H(A_X^{init} E) \tag{8}$$
$$= H(A'B) - H(AB) := \Delta H \tag{9}$$

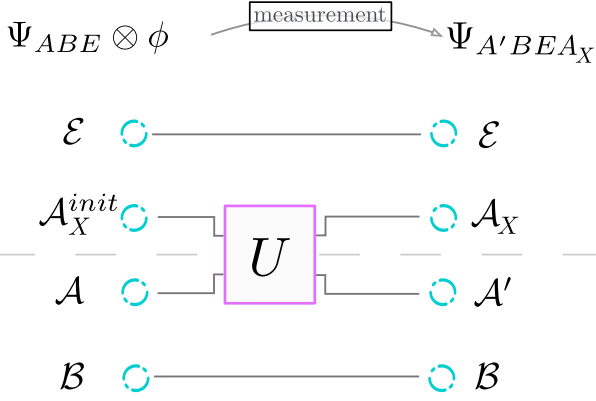

Fig. 1. **The Stinespring dilated perspective:** A key generating measurement $A_X$ is regarded from the perspective of a higher Hilbert space by including the action $\mathcal{A}_X \to \mathcal{A}_X^{init}$ on a classical memory (an abelian algebra). From this perspective the global state before and after the measurement is pure. In this case, due to the Araki-Lieb inequality (see Thm.2(c) of [5]), the entropy production $\Delta H = H(A_X|E)$ on the Eve-Memory partition equals the entropy change on the Alice-Bob partition. This allows us to bound $H(A_X|E)$ by measuring only on the Alice-Bob partition, i.e. without accessing Eve's system.

Therefore, we could restrict the values of $H(A_X|E)$ by only considering the transformation $\mathcal{AB} \to \mathcal{A'B}$. A short calculation (see [1–3]) shows that this transformation is described by the pinching channels, defined by

$$T_0[\rho] := (\Pi_0^{A_0} \otimes \mathbf{1})\rho(\Pi_0^{A_0} \otimes \mathbf{1}) + (\Pi_1^{A_0} \otimes \mathbf{1})\rho(\Pi_1^{A_0} \otimes \mathbf{1})$$
$$T_1[\rho] := (\Pi_0^{A_1} \otimes \mathbf{1})\rho(\Pi_0^{A_1} \otimes \mathbf{1}) + (\Pi_1^{A_1} \otimes \mathbf{1})\rho(\Pi_1^{A_1} \otimes \mathbf{1}). \tag{10}$$

Following the definitions of the pinching channels, we have the identities $T[\rho] = T \circ T[\rho] = T^*[\rho]$ and also $f(T[\rho]) = T[f(T[\rho])]$ when used within a functional calculus for a measurable function $f$. Hence, we can conclude that:

$$\begin{aligned} H(T[\rho]) - H(\rho) &= -\operatorname{tr}(T[\rho]\log(T[\rho])) + \operatorname{tr}(\rho\log(\rho)) \\ &= -\operatorname{tr}(\rho T^*[\log(T[\rho])]) + \operatorname{tr}(\rho\log(\rho)) \\ &= -\operatorname{tr}(\rho\log(T[\rho])) + \operatorname{tr}(\rho\log(\rho)) \\ &= D(\rho\|T[\rho]), \end{aligned} \tag{11}$$

where $D(\rho\|\sigma)$ denotes the relative entropy. Therefore, we can rewrite the convex mixture of conditional entropies as

$$\begin{aligned} &\lambda H(A_0|E) + (1-\lambda)H(A_1|E) \\ &= \lambda H(T_0[\rho_{AB}]) - \lambda H(\rho_{AB}) + (1-\lambda)H(T_1[\rho_{AB}]) - (1-\lambda)H(\rho_{AB}) \\ &= \lambda D(\rho_{AB}\|T_0[\rho_{AB}]) + (1-\lambda)D(\rho_{AB}\|T_1[\rho_{AB}]) \end{aligned} \tag{12}$$

Thus, we are able to bound Eve's conditional uncertainty on the measurement outcomes from any convex mixture of measurement bases by estimating the entropy production on the Alice-Bob system.

### B. Restriction to qubits

In this subsection, we will show that for the purpose of bounding $C^*(S)$, we could without loss of generality consider Alice-Bob system as a convex combination of two-qubits states.

A fundamental theorem on the algebra generated by a pair of two projections [6–9], which will be in our case $(\Pi_0^{A_0}, \Pi_0^{A_1})$ such as $(\Pi_0^{B_2}, \Pi_0^{B_3})$, states that we can always find a representation of the respective system of Alice and Bob that allows us to decompose the projectors into $2 \times 2$ blocks and a commuting rest.

In particular, let $K^A, L^A, K^B$ and $L^B$ be pairwise commuting projectors and let

$$Q(\theta) = \begin{pmatrix} \cos(\theta/2)^2 & \cos(\theta/2)\sin(\theta/2) \\ \cos(\theta/2)\sin(\theta/2) & \sin(\theta/2)^2 \end{pmatrix} \tag{13}$$

be a family of $2 \times 2$ projectors. Then, we can decompose

$$\Pi_0^{A_0} = \bigoplus_j \begin{pmatrix} 1 & 0 \\ 0 & 0 \end{pmatrix} \oplus K^A \text{ and } \Pi_0^{A_1} = \bigoplus_j Q(\varphi_j) \oplus L^A \tag{14}$$

$$\Pi_0^{B_2} = \bigoplus_l \begin{pmatrix} 1 & 0 \\ 0 & 0 \end{pmatrix} \oplus K^B \text{ and } \Pi_0^{B_3} = \bigoplus_l Q(\omega_l) \oplus L^B \tag{15}$$

where $\{\varphi_j\}$ and $\{\omega_l\}$ are families of angles that can be obtained from the spectra of $\sigma(\Pi_0^{A_0} + \Pi_0^{A_1})$ and $\sigma(\Pi_0^{B_2} + \Pi_0^{B_3})$ respectively (see [9] for details). The representations of the remaining projectors $\Pi_1^{A_0}, \Pi_1^{A_1}, \Pi_1^{B_2}$ and $\Pi_1^{B_3}$ follow from the identities

$$\Pi_1^{A_X} = \mathbf{1} - \Pi_0^{A_X} \text{ and } \Pi_1^{B_Y} = \mathbf{1} - \Pi_0^{B_Y}. \tag{16}$$

Furthermore, we can also always find channels $T_{block}^A, T_{block}^B$ and $T_{block}^{AB} := T_{block}^A \otimes T_{block}^B$ that diagonalise the entire systems of Alice and Bob into the corresponding blocks of (14) and (15). Applying these channels on an arbitrary state $\rho_{AB}$ will therefore give us a corresponding decomposition

$$T_{block}^{AB}[\rho_{AB}] = \bigoplus_{jl} \mu_{jl} \rho^{jl} \oplus \mu_{rest} \sigma_{rest}, \tag{17}$$

where $\sigma_{rest}$ denotes the projection of $\rho_{AB}$ into blocks that are commuting on either Alice's and/or Bob's side while the coefficients $\mu_{jl}$ can be obtained by renormalising $T_{block}^{AB}[\rho_{AB}]$ blockwise (see [4]). As the pinching channels $T_0$ and $T_1$ commute with $T_{block}^{AB}$ by construction and by using data processing, we can estimate

$$\lambda D(\rho_{AB} \| T_0[\rho_{AB}]) + (1-\lambda) D(\rho_{AB} \| T_1[\rho_{AB}])$$
$$\geq \lambda D(T_{block}^{AB}[\rho_{AB}] \| T_0 \circ T_{block}^{AB}[\rho_{AB}]) + (1-\lambda) D(T_{block}^{AB}[\rho_{AB}] \| T_1 \circ T_{block}^{AB}[\rho_{AB}]))$$
$$= \sum_{jl} \mu_{il} \left( \lambda D(\rho_{AB}^{jl} \| T_0[\rho_{AB}^{jl}]) + (1-\lambda) D(\rho_{AB}^{jl} \| T_1[\rho_{AB}^{jl}]) \right) + \mu_{rest} \left( \lambda D(\sigma_{rest} \| T_0[\sigma_{rest}]) + (1-\lambda) D(\sigma_{rest} \| T_1[\sigma_{rest}]) \right). \tag{18}$$

Relating back to the constraint of optimisation (7), we have $CHSH = T_{block}^{AB}[CHSH]$ which implies

$$\mathrm{tr}\left(\rho_{AB} CHSH\right) = S \Leftrightarrow \mathrm{tr}\left(\rho_{AB}^{jl} CHSH\right) = S^{jl} \text{ and } \sum_{jl} \mu_{jl} S^{jl} = S, \tag{19}$$

where we already took into account that $\sigma_{rest}$ will not play any role in the following optimisation as $|\mathrm{tr}(\sigma_{rest} CHSH)| \leq 2$ since the corresponding blocks commute on at least one side. Using the above relations, we can optimise every $\rho_{AB}^{jl}$ individually and reformulate (7) as

$$C^*(S) = \inf_{A_0,A_1,B_2,B_3} \inf_{\substack{\{\mu_{jl}\} \\ \sum_{jl} \mu_{jl} \leq 1}} \inf_{\substack{\{S^{jl}\} \\ \sum_{jl} \mu_{jl} S^{jl} = S}} \sum_{jl} \mu_{jl} \quad \begin{array}{l} \inf_{\rho^{jl}} \lambda D(\rho_{AB}^{jl} \| T_0[\rho_{AB}^{jl}]) + (1-\lambda) D(\rho_{AB}^{jl} \| T_1[\rho_{AB}^{jl}]) \\ \text{s.th.: } \langle CHSH \rangle_{\rho_{AB}^{jl}} = S^{jl} \end{array} \tag{20}$$

In the above we lower bound $\inf_{\rho^{jl}}[\ldots]$ by the function

$$C^*_{\mathbb{C}^{4\times 4}}(S) = \inf_{\varphi,\omega} \begin{array}{l} \inf_{\rho^{jl}} \lambda D(\rho_{AB}^{jl} \| T_0[\rho_{AB}^{jl}]) + (1-\lambda) D(\rho_{AB}^{jl} \| T_1[\rho_{AB}^{jl}]) \\ \text{s.th.: } \langle CHSH \rangle_{\rho_{AB}^{jl}} = S \end{array} \tag{21}$$

Since (21) is by construction independent of the specific angles of the underlying measurements $A_0, A_1, B_2, B_3$, we can reduce the optimisation over all measurements $A_0, A_1, B_2, B_3$ to an optimisation over all possible $\{\mu_{jl}\}$. Hence we can conclude the bound

$$C^*(S) \geq \inf_{A_0,A_1,B_2,B_3} \inf_{\substack{\{\mu_{jl}\} \\ \sum_{jl} \mu_{jl} \leq 1}} \inf_{\substack{\{S^{jl}\} \\ \sum_{jl} \mu_{jl} S^{jl} = S}} \sum_{jl} \mu_{jl} \ C^*_{\mathbb{C}^{4\times 4}}(S^{jl})$$
$$\geq \inf_{\mu} \int_{S'=2}^{2\sqrt{2}} \mu(dS') C^*_{\mathbb{C}^{4\times 4}}(S')$$
$$\text{s.th.: } \mu([2, 2\sqrt{2}]) \leq 1, \quad \mu \geq 0, \quad \int_{S'=2}^{2\sqrt{2}} \mu(dS') S' = S, \tag{22}$$

by estimating the optimisation over all $\{\mu_{jl}\}$ of arbitrary length via an integral over a positive sub-normalised measure $\mu$. The boundaries of the above integration originate from the simple observation that $S$ can not be larger than $2\sqrt{2}$ and that $C^*_{\mathbb{C}^{4\times 4}}(S^{jl}) = 0$ for all $S \leq 2$, which in fact also gives us the justification to ignore the $\sigma_{rest}$ contribution in (18).

At this point, we can write down an explicit matrix representation of the CHSH operator. For reasons that will be clear in the following steps, it is convenient to decompose the CHSH operator with respect to Bob's measurements as

$$CHSH = F_0 + b_x F_x + b_z F_z \tag{23}$$

with

$$b_x = \sin(\omega) \qquad b_z = \cos(\omega)$$

$$F_0 = \begin{pmatrix} \cos(\varphi) - 1 & 0 & -\sin(\varphi) & 0 \\ 0 & 1 - \cos(\varphi) & 0 & \sin(\varphi) \\ -\sin(\varphi) & 0 & 1 - \cos(\varphi) & 0 \\ 0 & \sin(\varphi) & 0 & \cos(\varphi) - 1 \end{pmatrix}$$

$$F_x = \begin{pmatrix} 0 & 2\cos^2\left(\frac{\varphi}{2}\right) & 0 & -\sin(\varphi) \\ 2\cos^2\left(\frac{\varphi}{2}\right) & 0 & -\sin(\varphi) & 0 \\ 0 & -\sin(\varphi) & 0 & -2\cos^2\left(\frac{\varphi}{2}\right) \\ -\sin(\varphi) & 0 & -2\cos^2\left(\frac{\varphi}{2}\right) & 0 \end{pmatrix}$$

$$F_z = \begin{pmatrix} -\cos(\varphi) - 1 & 0 & \sin(\varphi) & 0 \\ 0 & \cos(\varphi) + 1 & 0 & -\sin(\varphi) \\ \sin(\varphi) & 0 & \cos(\varphi) + 1 & 0 \\ 0 & -\sin(\varphi) & 0 & -\cos(\varphi) - 1 \end{pmatrix} \tag{24}$$

and reformulate the optimisation over Bob's angle $\omega$ as an optimisation over two parameters $b_x, b_z$ fulfilling the constraint $b_x^2 + b_z^2 = 1$. Hence our optimisation problem for a single two-qubit state reads

$$C^*_{\mathbb{C}^{4\times 4}}(S) := \inf_{\varphi \in [0, \pi/2]} \inf_{\substack{(b_x, b_z) \\ b_x^2 + b_z^2 = 1}} \begin{array}{l} \inf_\rho \lambda D(\rho \| T_0[\rho]) + (1-\lambda) D(\rho \| T_1[\rho]) \\ \text{s.th.: } \langle F_0 \rangle_\rho + b_x \langle F_x \rangle_\rho + b_z \langle F_z \rangle_\rho = S \end{array}, \tag{25}$$

where we will, with a slight abuse of notation, denote the pinching channels for the reduced $2 \times 2$ measurements as $T_0$ and $T_1$ as well.

### C. Estimating the relative entropy via the trace norm (Pinsker++)

In this subsection, we will show that a lower bound of $C^*_{\mathbb{C}^{4\times 4}}(S)$ is related to an optimisation with its objective function given by a combination of trace norms. As such, this optimisation can be efficiently bounded as shown in the following subsections. Here, we employ a refined version of Pinsker's inequality provided in Theorem 1 (also see [10]): for a pinching channel with two-outcomes we can lower bound the relative entropy between its input and output by their trace norm via

$$D(\rho \| T_X[\rho]) \geq \log(2) - h_2\left(\frac{1}{2} - \frac{1}{2} \|\rho - T_X[\rho]\|_1\right), \tag{26}$$

where $h_2(p) = -p \log p - (1-p) \log(1-p)$ denotes the binary entropy function. Using the above inequality and the concavity of the binary entropy, we could write

$$\lambda D(\rho \| T_0[\rho]) + (1-\lambda) D(\rho \| T_1[\rho]) \tag{27}$$

$$\geq \log(2) - \lambda h_2\left(\frac{1}{2} - \frac{1}{2} \|\rho - T_0[\rho]\|_1\right) - (1-\lambda) h_2\left(\frac{1}{2} - \frac{1}{2} \|\rho - T_1[\rho]\|_1\right)$$

$$\geq \log(2) - h_2\left(\frac{1}{2} - \frac{1}{2}\left(\lambda \|\rho - T_0[\rho]\|_1 + (1-\lambda) \|\rho - T_1[\rho]\|_1\right)\right)$$

which is a monotonous function of the sum of the trace norms. Hence, we can employ the inequality

$$C^*_{\mathbb{C}^{4\times 4}}(S) \geq \log(2) - h_2\left(\frac{1}{2} - \frac{1}{2}t\right) \tag{28}$$

for any $t \leq t^*(S)$ that gives a lower bound on the optimisation

$$t^*(S) := \boxed{\inf_{\varphi \in [0,\pi/2]} \boxed{\inf_{\substack{(b_x,b_z) \\ b_x^2+b_z^2=1}} \boxed{\begin{array}{l} \inf_\rho \quad \lambda \|\rho - T_0[\rho]\|_1 + (1-\lambda)\|\rho - T_1[\rho]\|_1 \\ \text{s.th.:} \quad \langle F_0 \rangle_\rho + b_x \langle F_x \rangle_\rho + b_z \langle F_z \rangle_\rho = S \end{array}}}} \tag{29}$$

The boxes in (29) highlight different stages in the optimisation that will be addressed individually in the following three subsections.

### D. Optimisation of $\rho$ for fixed angles: reformulation as SDP

In this subsection, we will formulate the minimisation over $\rho$ (black box in (29)) as a SDP, which can be solved using existing numerical algorithms. To this end, we will use a well-known relation that the trace norm of a quadratic matrix $V$ can be represented by the minimisation (refer to [11])

$$\|V\|_1 = \inf_{K,L} \frac{1}{2} \operatorname{tr}(K+L) \quad \text{s.th.:} \begin{pmatrix} K & V \\ V^\dagger & L \end{pmatrix} \geq 0. \tag{30}$$

over additional matrices $K$ and $L$. We can explicitly expand the channels $T_X$ in terms of the measurement projectors, which can be written as

$$\rho - T_1(\rho) = \rho Q(\varphi) + Q(\varphi)\rho - 2Q(\varphi)\rho Q(\varphi) \text{ and } \rho - T_0(\rho) = \rho Q(0) + Q(0)\rho - 2Q(0)\rho Q(0). \tag{31}$$

Hence, we can rewrite the minimisation over $\rho$ as the following SDP:

$$\boxed{\begin{array}{l} \inf_\rho \quad \lambda \|\rho - T_0[\rho]\|_1 + (1-\lambda)\|\rho - T_1[\rho]\|_1 \\ \text{s.th.:} \quad \langle F_0 \rangle_\rho + b_x \langle F_x \rangle_\rho + b_z \langle F_z \rangle_\rho = S \end{array}} \tag{32}$$

$$= \inf_{\substack{\rho \\ K_0,L_0,K_1,L_1}} \lambda \operatorname{tr}(K_0 + L_0) + (1-\lambda)\operatorname{tr}(K_1 + L_1)$$

$$\text{s.th.:} \begin{pmatrix} K_0 & \rho Q(0) + Q(0)\rho - 2Q(0)\rho Q(0) \\ \rho Q(0) + Q(0)\rho - 2Q(0)\rho Q(0) & L_0 \end{pmatrix} \geq 0$$

$$\begin{pmatrix} K_1 & \rho Q(\varphi) + Q(\varphi)\rho - 2Q(\varphi)\rho Q(\varphi) \\ \rho Q(\varphi) + Q(\varphi)\rho - 2Q(\varphi)\rho Q(\varphi) & L_1 \end{pmatrix} \geq 0$$

$$\operatorname{tr}\left(\rho(F_0 + b_x F_x + b_z F_z)\right) = S. \tag{33}$$

### E. Optimisation of Bob's angle: an outer approximation to $b_x, b_z$

The minimisation over Bob's measurement angle (the magenta box in (29)) was formulated as an optimisation over parameters $(b_x, b_z)$ from a set

$$\mathcal{K} = \{(b_x, b_z) \mid b_x^2 + b_z^2 = 1\}, \tag{34}$$

which can be interpreted as boundary of a circle with unit radius. Since the objective function of (29) does not explicitly depends on $(b_x, b_z)$, we could move this minimisation directly into the constraints and write

$$
\boxed{\inf_{\substack{b_x, b_z \\ b_x^2 + b_z^2 = 1}} \boxed{\begin{aligned} \inf_\rho \quad & \lambda \left\| \rho - T_0[\rho] \right\|_1 + (1-\lambda) \left\| \rho - T_1[\rho] \right\|_1 \\ \text{s.th.:} \quad & \langle F_0 \rangle_\rho + b_x \langle F_x \rangle_\rho + b_z \langle F_z \rangle_\rho = S \end{aligned}}} \tag{35}
$$

$$
= \begin{aligned} \inf_\rho \quad & \lambda \left\| \rho - T_0[\rho] \right\|_1 + (1-\lambda) \left\| \rho - T_1[\rho] \right\|_1 \\ \text{s.th.:} \quad & \exists (b_x, b_z) \in \mathcal{K}: \quad \langle F_0 \rangle_\rho + b_x \langle F_x \rangle_\rho + b_z \langle F_z \rangle_\rho = S \end{aligned} \tag{36}
$$

$$
= \begin{aligned} \inf_\rho \quad & \lambda \left\| \rho - T_0[\rho] \right\|_1 + (1-\lambda) \left\| \rho - T_1[\rho] \right\|_1 \\ \text{s.th.:} \quad & \sup_{(b_x, b_z) \in \mathcal{K}} \langle F_0 \rangle_\rho + b_x \langle F_x \rangle_\rho + b_z \langle F_z \rangle_\rho \geq S. \end{aligned} \tag{37}
$$

Clearly, the above problem is no longer a convex optimisation problem but there exists a convex relaxation by replacing $\mathcal{K}$ with any convex set $\mathcal{P} \supseteq \mathcal{K}$. We will replace $\mathcal{K}$ by a compact convex polytope $\mathcal{P}$ with finite set of vertices $(p_x^l, p_z^l)$. Since the individual constraints in (35) are linear in $b_x, b_z$, it suffices to only consider points on these vertices. As such, we can efficiently lower bound (35) via

$$
(35) \geq \begin{aligned} \inf_\rho \quad & \lambda \left\| \rho - T_0[\rho] \right\|_1 + (1-\lambda) \left\| \rho - T_1[\rho] \right\|_1 \\ \text{s.th.:} \quad & \sup_{(b_x, b_z) \in \mathcal{P}} \langle F_0 \rangle_\rho + b_x \langle F_x \rangle_\rho + b_z \langle F_z \rangle_\rho \geq S \end{aligned} \geq \begin{aligned} \inf_l \inf_\rho \quad & \lambda \left\| \rho - T_0[\rho] \right\|_1 + (1-\lambda) \left\| \rho - T_1[\rho] \right\|_1 \\ \text{s.th.:} \quad & \langle F_0 \rangle_\rho + p_x^l \langle F_x \rangle_\rho + p_z^l \langle F_z \rangle_\rho \geq S \end{aligned} \tag{38}
$$

The above structure can now be incorporated in a simple algorithm (illustrated in Fig. 2) that computes the optimisation of Bob's measurements via a sequence of lower bounds (also see [12, 13] where a similar method was used for 3-dimensional problems). The algorithm is given by:

(i) Start by initiating $\mathcal{P}$ as a rectangle that contains $\mathcal{K}$.

(ii) Use (32) to compute the value of (38) for every vertices $(p_x^1, p_z^1) \ldots (p_x^n, p_z^n)$ of $\mathcal{P}$.

(iii) Pick the first vertex $b_{min}$ that attains the minimum in (38) and refine $\mathcal{P}$ by introducing a new edge to $\mathcal{P}$ that is tangential to $\mathcal{K}$. This will "cut off" $b_{min}$ and create two new vertices (see Fig.2). Proceed to (ii).

Repeating $(ii)$ and $(iii)$ will give a progressive refinement of $\mathcal{P}$ that approaches the boundary of $\mathcal{K}$ from the outside at the vicinity of those points $(b_x, b_z)$ that realise the minimum of (29). It turns out that, this algorithm only requires a few iterations to achieve a very high precision for our above-mentioned problem.

### F. Optimisation of Alice's angle: $\boxed{\text{an } \varepsilon\text{-net for } \varphi}$

For the representation given by equation (14), Alice's measurement on her qubit is fully described by the angle $\varphi$, which could take values in the interval $[0, \pi/2]$. Consequently, the minimisation over Alice's measurement angle (cyan box in (29)) is not an SDP.

In this subsection, we will show how one could obtain a bound on the minimisation over Alice's measurement angle by employing an $\varepsilon$-net. The basic idea of the $\varepsilon$-net is as follows: we split the interval $[0, \pi/2]$ into smaller segments of size $2\varepsilon_0$ each centred around a discretised range-point $\varphi_i$. We then evaluate (29) (i.e. everything in the magenta box) on each discretised range-point and subtract a pessimistic error $\Delta(\varepsilon_0, \varphi)$, which bounds how much would the value of (29) have changed when evaluated on any other $\varphi$ within the segment. Taking the minimal value over all segments will then give us a lower bound on the whole optimisation (29)(cyan box).

Refining those segments into even smaller ones will give tighter lower bounds. In order to keep the total computational cost low, we successively refine only the segment that gives the minimum value for a given round of the procedure. By doing so, we arrive in a situation (see Fig.3) where we only have to improve our refinement around a single point corresponding to the global minimum, i.e. the single $\varphi$ that achieves $C^*_{\mathbb{C}^{4 \times 4}}(S)$.

In order to bound the error of each segment, we will need to investigate the relation between the entire optimisation and Alice's measurement angle $\varphi$. Varying $\varphi$ (see (14)) will affect the two inner optimisations of (29) in two ways: On

7

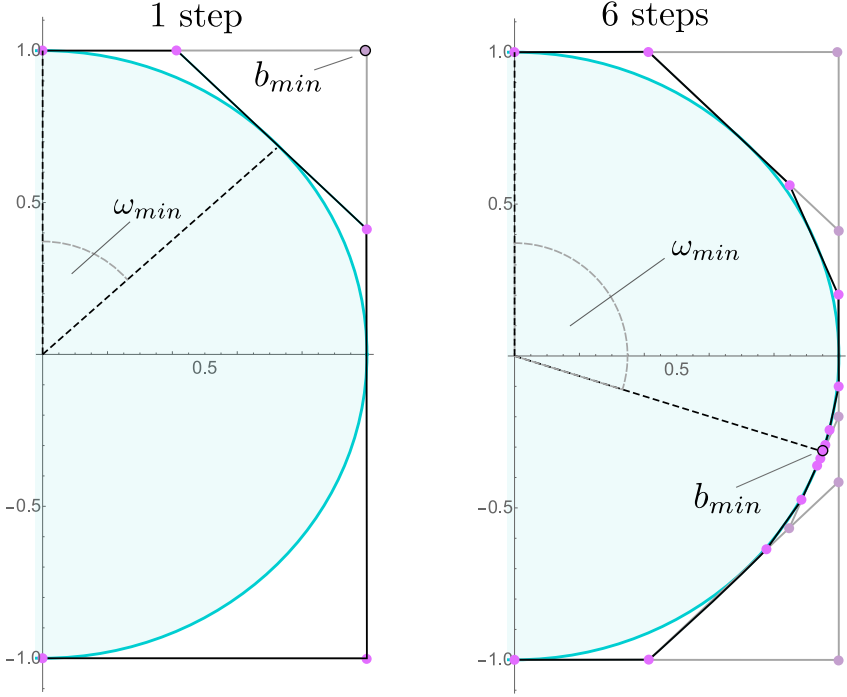

Fig. 2. **Minimising $\omega$:** Iterative refinement of $\mathcal{P}$ (magenta vertices, black edges) to identify the optimal $\omega$ in $\mathcal{K}$ (cyan semicircle). The parameters considered in the diagram is given by $\lambda = 0.2$, $S = 2.4$ and $\varphi = 1.2$. In every iteration of our algorithm, we identify $b_{min}$ (grey circles), the vertex of $\mathcal{P}$ that gives the minimum value of optimisation (38), and remove it by introducing a new edge. In this example, six iterations are sufficient to achieve a target precision of $10^{-5}$.

one hand, since the objective function depends on $\varphi$ via the $T_1(\rho)$, it will change with varying $\varphi$. On the other hand, the feasible set of possible $\rho_{AB}$ will change with varying $\varphi$ since the CHSH operator, i.e. $F_0, F_x$ and $F_z$, depends on $\varphi$ as well.

We will now address the dependence of the CHSH operator on $\varphi$. Without loss of generality we could have reformulated (29) by considering all states that could achieve a Bell violation *greater* than or equals the observed CHSH value $S$ as it is a valid relaxation of the optimisation problem. This implies that for fixed $\varphi$ we could also perform our optimisation over the set

$$\mathcal{S}_\varphi^S := \{\rho | \langle CHSH|_\varphi \rangle_\rho \geq S\}. \tag{39}$$

When we change $\varphi$ to $\varphi + \varepsilon$, on a segment specified by $|\varepsilon| \leq \varepsilon_0$, we have to consider a different set of states $\mathcal{S}_{\varphi+\varepsilon}^s$. Hence, the optimal solution on the whole segment around $\varphi$ will be attained on a state from the set

$$\overline{\mathcal{S}}_{\varepsilon_0 \varphi}^S := \bigcup_{|\varepsilon| \leq \varepsilon_0} \mathcal{S}_{\varphi+\varepsilon}^S \tag{40}$$

We consider the norm of the difference of the two CHSH operators and we have:

$$\|CHSH|_\varphi - CHSH|_{\varphi-\varepsilon}\|_\infty$$
$$= \|(F_0 + b_x F_x + b_z F_z)|_\varphi - (F_0 + b_x F_x + b_z F_z)|_{\varphi-\varepsilon}\|_\infty$$
$$= 2\|(Q(\varphi) - Q(\varphi - \varepsilon)) \otimes (C^{B_2} + C^{B_3})\|_\infty$$
$$\leq 4\|Q(\varphi) - Q(\varphi - \varepsilon)\|_\infty$$
$$\leq 4 \max_{\varphi, |\varepsilon| \leq \varepsilon_0} \|Q(\varphi) - Q(\varphi - \varepsilon)\|_\infty \tag{41}$$

$$= 4 \max_{\varphi, |\varepsilon| \leq \varepsilon_0} \left\| \begin{pmatrix} \cos(\frac{\varphi}{2})^2 - \cos(\frac{\varphi-\varepsilon}{2})^2 & \cos(\frac{\varphi}{2})\sin(\frac{\varphi}{2}) - \cos(\frac{\varphi-\varepsilon}{2})\sin(\frac{\varphi-\varepsilon}{2}) \\ \cos(\frac{\varphi}{2})\sin(\frac{\varphi}{2}) - \cos(\frac{\varphi-\varepsilon}{2})\sin(\frac{\varphi-\varepsilon}{2}) & \sin(\frac{\varphi}{2})^2 - \sin(\frac{\varphi-\varepsilon}{2})^2 \end{pmatrix} \right\|_\infty \tag{42}$$

$$= 4 \max_{|\varepsilon| \leq \varepsilon_0} \sqrt{\frac{1}{2} - \frac{1}{2}\cos(\varepsilon)} \tag{43}$$

$$\leq 2\varepsilon_0, \tag{44}$$

where we assumed that $b_x^2 + b_z^2 = 1$ holds up to a negligible deviation as sufficient iterations are taken in the outer approximation shown in Fig. 2 (otherwise, adding a further factor $\approx 1 + \epsilon$ will be required). From the above relation, we conclude that every state $\rho$ that could attain a CHSH value $S$ for measurements with the measurement angle $\varphi + \varepsilon$ will attain a CHSH value of at least $S - 2\varepsilon_0$ with the measurements angle $\varphi$. Hence, we have the inclusion relation:

$$\overline{\mathcal{S}}_{\varepsilon_0 \varphi}^S \subseteq \mathcal{S}_\varphi^{S-2\varepsilon_0}. \tag{45}$$

8

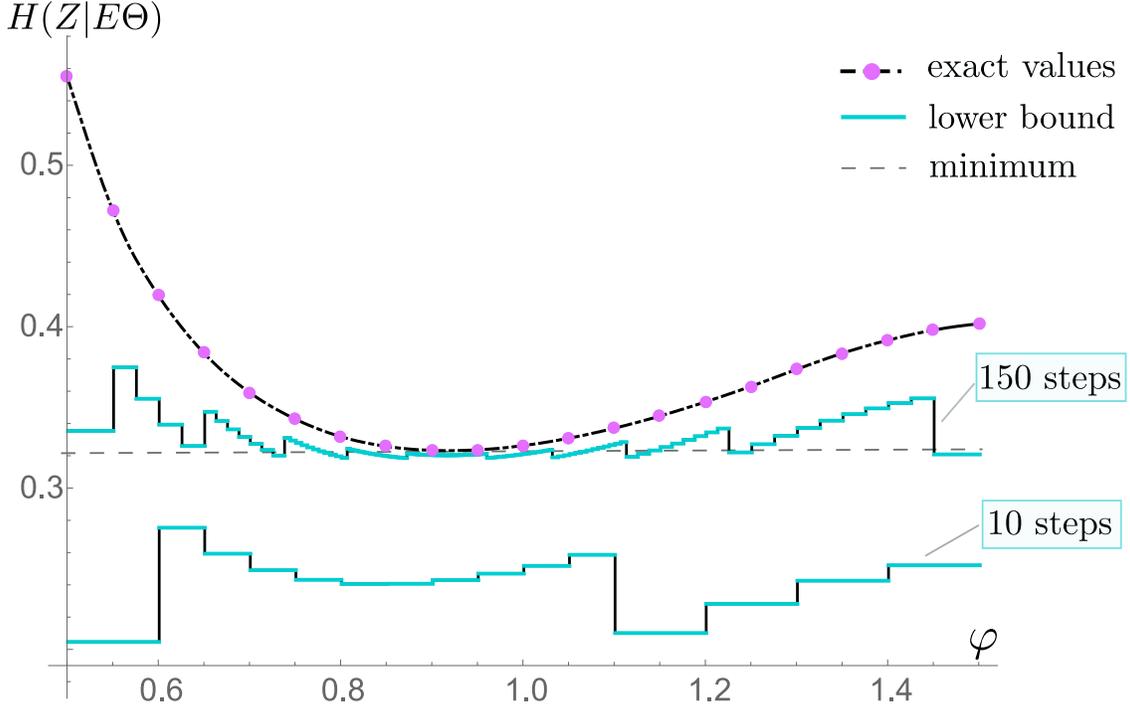

Fig. 3. **Adaptive refinement of the $\varepsilon$-net** Lower bounds on $H(Z|E\theta)$, by evaluating (28) on segments. (For comparison) The magenta coloured points give the exact value of (29) for a discretised range of different $\varphi$. (The actual bound) The cyan coloured line gives a pessimistic lower bound (not assuming continuity) valid for a segment of length $\varepsilon_0$ centred around a specific $\varphi$. Fixing the width of a segment, $\varepsilon_0$, to be fine enough will give a tighter lower bound. In order to perform the computation efficiently, we start with a rough discretised range with a large segment size. We then identify and refine the segment with the lowest estimate on the target function $H(Z|E\theta)$ and we repeat this procedure.

which means that we can perform optimisation (29) by replacing $S$ with $S - 2\varepsilon_0$ on each segment of size $2\varepsilon_0$ (see Fig. 4) to obtain a reliable lower bound.

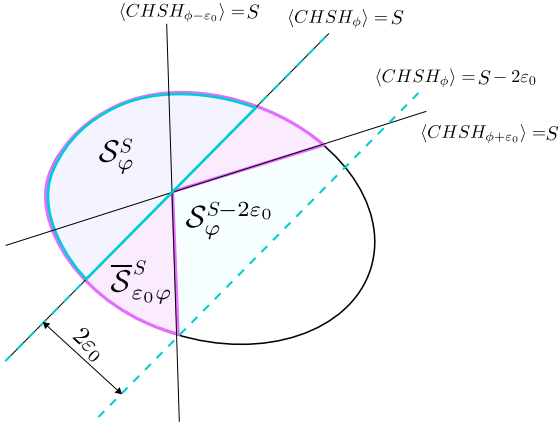

Fig. 4. **Feasible sets:** The optimisation for a fixed $\varphi$ runs over a set $\mathcal{S}_\varphi^S$, i.e. all states with CHSH values larger than or equals to $S$. Varying $\varphi$ by not more than $\varepsilon_0$ will not change the CHSH value by more than $2\varepsilon_0$. Therefore, the set $\mathcal{S}_\varphi^{S-2\varepsilon_0}$ includes all states that can attain CHSH value larger than or equals to $S$ for some angle in the interval $(\varphi - \varepsilon_0, \varphi + \varepsilon_0)$.

We are now left with assessing the change in the objective of (29) introduced by varying $\varphi$, i.e. the error of the functional. Using Hölder's inequality, we can dualise the trace norm of an operator $X$ as

$$\|X\|_1 = \sup_{Y:\|Y\|_\infty = 1} \operatorname{tr}(YX) \tag{46}$$

and for suitably chosen $M, N$ with $\|M\|_\infty = \|N\|_\infty = 1$, we can rewrite the objective of (29) as

$$f_{M,N}(\varphi, \rho) := \langle \lambda\{M, Q(0)\} - \lambda Q(0) M Q(0) + (1-\lambda)\{N, Q(\varphi)\} - (1-\lambda) Q(\varphi) N Q(\varphi)\rangle_\rho \tag{47}$$

9

Here, we could bound the following expression by:

$$|f_{M,N}(\varphi,\rho) - f_{M,N}(\varphi+\varepsilon,\rho)|$$
$$\leq (1-\lambda) \sup_{N,\varphi,\varepsilon} \left\| \{N, Q(\varphi)\} - Q(\varphi)NQ(\varphi) - \{N, Q(\varphi+\varepsilon)\} + Q(\varphi+\varepsilon)NQ(\varphi+\varepsilon) \right\|_\infty$$
$$\leq (1-\lambda)4\varepsilon_0 \qquad (48)$$

which follows after a short computation from the central estimate $\|(Q(\varphi) - Q(\varphi+\varepsilon))\|_\infty \leq \varepsilon_0/2$, which was already computed in (44).

Thus, by combining (44) and (48), we can estimate the maximal error on a segment around some fixed $\varphi$. Applying this estimate and the technique described above to (29), finally gives

$$t^*(S) = \inf_{\varphi \in [0,\pi/2]} \inf_{\substack{(b_x, b_z) \\ b_x^2 + b_z^2 = 1}} \begin{array}{l} \inf_\rho \quad \lambda \|\rho - T_0[\rho]\|_1 + (1-\lambda) \|\rho - T_1[\rho]\|_1 \\ \text{s.th.:} \quad \langle F_0 \rangle_\rho + b_x \langle F_x \rangle_\rho + b_z \langle F_z \rangle_\rho = S \end{array}$$

$$\geq \inf_{\varphi_k \in \mathcal{I}} \inf_l \begin{array}{l} \inf_\rho \quad \lambda \|\rho - T_0[\rho]\|_1 + (1-\lambda) \|\rho - T_1[\rho]\|_1 - 2(1-\lambda)\varepsilon_0^k \\ \text{s.th.:} \quad \langle F_0 \rangle_\rho + b_x \langle F_x \rangle_\rho + b_z \langle F_z \rangle_\rho = S - 2\varepsilon_0^k \end{array} \qquad (49)$$

where $\varepsilon_0^k$ denotes the according segment width and $\varphi_k \in \mathcal{I}$ denote the angles which form the discretised range as described above and in Fig. 3.

### G. Worst case convex hull over two-qubit strategies:

By employing the techniques described in the previous subsections, we are able to compute the two-qubit function $C^*_{\mathbb{C}^{4\times 4}}(S)$ point-wise up to arbitrary precision. The only remaining step for computing the desired bound $C^*(S)$ is to estimate (22). We will tackle this problem by defining a convex function $\overline{C}(S)$ that is smaller or equals to $C^*_{\mathbb{C}^{4\times 4}}(S)$ for all $S \in (2, 2\sqrt{2})$, (see Fig. 5). Given such a function $\overline{C}(S)$, we can conclude that

$$C^*(S) = \inf_{\mu: \langle S' \rangle_\mu = S} \int_{S'=2}^{2\sqrt{2}} \mu(dS') C^*_{\mathbb{C}^{4\times 4}}(S')$$
$$\geq \inf_{\mu: \langle S' \rangle_\mu = S} \int_{S'=2}^{2\sqrt{2}} \mu(dS') \overline{C}(S')$$
$$= \inf_{\mu: \langle S' \rangle_\mu = S} \overline{C}\left( \int_{S'=2}^{2\sqrt{2}} \mu(dS') S' \right) = \overline{C}(S). \qquad (50)$$

At this point, we note that the analytical result of [4], which is a special case of our computation for $\lambda = 1$, shows that $C^*_{\mathbb{C}^{4\times 4}}(S)$ is convex. This means that we can set $\overline{C}(S) = C^*_{\mathbb{C}^{4\times 4}}(S)$ in the case of $\lambda = 1$. The numerical evaluation of $C^*_{\mathbb{C}^{4\times 4}}(S)$ suggest the same conclusion might actually hold for all $\lambda \in (0,1)$. However, we could not prove this analytically and therefore, the following steps is necessary.

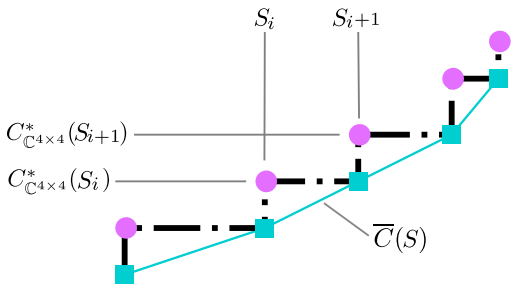

Fig. 5. **Convex bound:** given the value of $C^*_{\mathbb{C}^{4\times 4}}(S)$ for a discretised range of values $\cdots \leq S_i \leq S_{i+1} \leq \ldots$ (magenta points) allows us to estimate $C^*_{\mathbb{C}^{4\times 4}}(S)$ from below by taking the floor on every interval (dot-dashed black line). The extreme points of this graph are a subset of the points marked by cyan squares. From this we can easily compute the lower convex estimate $\overline{C}(S)$ (cyan line).

It is straight forward to see from the definition of (25) that $C^*_{\mathbb{C}^{4\times 4}}(S)$ is a monotonous non-decreasing function, i.e. we have $C^*_{\mathbb{C}^{4\times 4}}(S_1) \leq C^*_{\mathbb{C}^{4\times 4}}(S_2)$ for $S_1 \leq S_2$, since increasing $S$ will impose more restrictions on Eve leaving her with a smaller

set of possible attacks. The evaluation of $C^*_{\mathbb{C}^{4\times 4}}(S)$ for a discretised range of values $S_0 = 2 \leq S_1 \leq S_2 \leq \cdots \leq S_N = 2\sqrt{2}$ will therefore enable us to estimate $C^*_{\mathbb{C}^{4\times 4}}(S)$ from below for all values of $S$ by taking the floor

$$\lfloor C^*_{\mathbb{C}^{4\times 4}}(S) \rfloor := C^*_{\mathbb{C}^{4\times 4}}(S_i) \qquad \forall S \in [S_i, S_{i+1}) \tag{51}$$

on every interval $[S_i, S_{i+1})$. Since this function is now defined on every $S$ and is never larger than $C^*_{\mathbb{C}^{4\times 4}}(S)$ itself, computing its lower convex hull by applying a Legendre transformation twice will give us a valid convex estimate $\overline{C}(S)$ (see Fig. 5).

### H. A conjectured alternative proof

We observed from some numerical exploration that the states saturating optimisation (7) for $\lambda = 1/2$ appear to have the following two properties:

1. The reduced state on $\mathcal{AB}$ is Bell-diagonal with only two nonzero eigenvalues.

2. The states satisfy $P_g(A_0|E) = P_g(A_1|E)$, where $P_g(A_X|E)$ is Eve's maximum probability of guessing the outcome of measurement $A_X$.

Letting $\rho_{E|A_X}$ denote Eve's state conditioned on the outcome $A_X$, the first property would imply $F(\rho_{E|A_X=0}, \rho_{E|A_X=1})^2 = 1 - d(\rho_{E|A_X=0}, \rho_{E|A_X=1})^2$, where $F$ is the root-fidelity and $d$ is the trace distance. (This is because Eve's states would then be supported on a common qubit subspace, and this equality holds for qubit states when the states have the same eigenvalues [14], which is indeed the case here [4].) Using the fact that we can assume $A_X$ to be uniform without loss of generality, it would then follow from the bound in [15] and the relation $d(\rho_{E|A_X=0}, \rho_{E|A_X=1}) = 2P_g(A_X|E) - 1$ that

$$H(A_X|E) \geq \log(2) - h_2\left(\frac{1 - F(\rho_{E|A_X=0}, \rho_{E|A_X=1})}{2}\right) = \log(2) - h_2\left(\frac{1 - \sqrt{4P_g(A_X|E)(1 - P_g(A_X|E))}}{2}\right). \tag{52}$$

Applying the second property would then imply[1]

$$\frac{1}{2}H(A_0|E) + \frac{1}{2}H(A_1|E) \geq \log(2) - h_2\left(\frac{1 - \sqrt{4P_g(A_X|E)(1 - P_g(A_X|E))}}{2}\right) \text{ for either } X$$

$$= \log(2) - h_2\left(\frac{1 - \sqrt{4P_g^{\text{avg}}(1 - P_g^{\text{avg}})}}{2}\right), \text{ where } P_g^{\text{avg}} = \frac{1}{2}P_g(A_0|E) + \frac{1}{2}P_g(A_1|E). \tag{53}$$

The key point of this reduction is that the maximum value of $P_g^{\text{avg}}$ (subject to a constraint on the CHSH value) can be bounded by SDP methods [16–18]. We computed the corresponding bounds, and found that they matched closely with the results obtained above. Hence if it could be proven that the two properties listed above indeed hold for the states saturating the optimisation in (7), this would provide an alternative approach to deriving our results. The benefit of such an approach is that it would be substantially faster, as it does not rely on computing an $\varepsilon$-net of points.

We remark that for the case $\lambda = 1$ (i.e. simply bounding the $H(A_0|E)$ term alone), we can prove that the first property holds, and thus Eq. (52) must hold in that scenario. In that case, we find that substituting the known closed-form bound on $P_g(A_0|E)$ in terms of the CHSH value [19] yields exactly the closed-form bound on $H(A_0|E)$ derived in Ref. [4]. This can be viewed as an alternative method to derive the latter bound, and it may be of interest to study how this approach could be generalised.

### I. Approaches without the qubit reduction

Our analysis thus far relied on reducing the analysis to a qubit scenario. In principle, however, there are some approaches which could be used to tackle the optimisation (7) without this reduction, and we shall now briefly outline these possibilities. These approaches could potentially allow for more general applicability of our methods.

---

[1] If the bound (52) were convex with respect to $P_g(A_X|E)$, the second property would not be required; unfortunately, the bound is instead concave, so instead this approach only yields the desired final result *if and only if* the second property holds.

The first approach consists of noting that the trace norm terms in the optimisation (29) can be rewritten using

$$\|\rho - T[\rho]\|_1 = \sup_{-\mathbf{1} \leq M \leq \mathbf{1}} \text{tr}(M(\rho - T[\rho])) = \sup_P \frac{1}{\|P\|_\infty} \text{tr}(P(\rho - T[\rho])) = \sup_P \frac{1}{\|P\|_\infty} \text{tr}((P - T[P])\rho), \quad (54)$$

where $\sup_P$ is taken over all Hermitian $P$, and for the last equality we used the fact that pinching channels are self-adjoint. Therefore, any choice of $P$ yields a valid lower bound. In particular, if we choose $P$ to be a (noncommutative) polynomial function of the measurement operators, then lower-bounding $\text{tr}((P - T[P])\rho)$ is precisely a noncommutative polynomial optimisation of the type studied in [18], which can be solved by SDP methods *without* reducing the analysis to qubits. An advantage of this approach is that we can impose the full output statistics as constraints (similar to [1, 16, 17]) instead of just the CHSH value alone. As for the $\|P\|_\infty$ term, it can be upper-bounded separately by noting that $\|P\|_\infty = \sup_\sigma |\text{tr}(P\sigma)| = \max\{\sup_\sigma \text{tr}(P\sigma), \sup_\sigma \text{tr}(-P\sigma)\}$, where $\sup_\sigma$ is taken over all normalised density operators $\sigma$. This is also of the right form to be bounded using the SDP method described in [18]. If $P$ is chosen to be a polynomial of low degree, the corresponding SDPs are quite tractable and can be solved quickly (in particular, they are substantially simpler than the ones described in [1]).

Unfortunately, when we applied this method to the CHSH scenario, we were unable to find a choice of $P$ that was sufficiently robust to noise for practical purposes. While we did find that it can yield a tight bound at maximum CHSH value, the bound quickly drops to zero once noise is added. It is not currently clear whether this is an inherent limitation of this approach, or whether some better choice of $P$ might overcome this difficulty. (A possible cause of this suboptimality could be the fact that since the optimisations for $\text{tr}((P - T[P])\rho)$ and $\|P\|_\infty$ were solved separately, this "decouples" them from each other, resulting in worse bounds than would be obtained from tackling (54) directly. The choice of $P$ that certifies a tight bound at maximum CHSH violation was obtained by exploiting the fact that the states and measurements which attain maximum CHSH value are essentially unique, up to trivial local operations. Therefore, we were able to find an explicit operator $P$ that saturates the variational characterisation in Eq. (54) and write it as a polynomial of those measurement operators. However, some care was still needed to choose the polynomial in such a way that the bound on $\|P\|_\infty$ is tight.)

A second approach is based on the method developed in [1]. To briefly summarise, the method described in that work essentially yields some possible choices of vector $\vec{\mu}$ and operator polynomials $K_0, K_1$ such that

$$H(A_X|E) \geq \vec{\mu} \cdot \vec{\gamma} - \log\langle K_X \rangle, \quad (55)$$

where $\vec{\gamma}$ is the list of all statistics estimated in the experiment (which could be, for instance, just the CHSH value alone but could also be all output statistics instead). We now note that for any $\alpha > 0$, we have $\log x \leq x/\alpha + \log \alpha - 1$ (this is simply the tangent line to $\log x$ at the point $x = \alpha$). Therefore, for any $\alpha_0, \alpha_1$ we would have

$$\lambda H(A_0|E) + (1-\lambda)H(A_1|E) \geq \lambda(\vec{\mu} \cdot \vec{\gamma} - \log\langle K_0\rangle) + (1-\lambda)(\vec{\mu} \cdot \vec{\gamma} - \log\langle K_1\rangle) \quad (56)$$

$$\geq \vec{\mu} \cdot \vec{\gamma} - \left\langle \frac{\lambda}{\alpha_0} K_0 + \frac{1-\lambda}{\alpha_1} K_1 \right\rangle - \lambda \log \alpha_0 - (1-\lambda)\log \alpha_1 + 1. \quad (57)$$

In this final expression, $\langle (\lambda/\alpha_0)K_0 + ((1-\lambda)/\alpha_1)K_1 \rangle$ is again an operator polynomial of a form which can be bounded using the SDPs in [18] without requiring the qubit reduction. By optimising over choices of the two tangent points $\alpha_X$ (as well as the vector $\vec{\mu}$ and operator polynomials $K_X$), we were able to compute some bounds on the main optimisation (7), imposing constraints based on the full output statistics. However, we found that while the resulting bounds are indeed better than the bounds on $H(A_0|E)$ alone, the improvement was fairly small (in particular, the resulting curves were worse than those shown in the main text). Hence we were also unable to use this approach to obtain tight bounds in this situation, though in contrast to the first approach above, it performs well at moderate noise values rather than low noise values. Again, it is not currently clear whether it might be possible to improve on this finding with a better choice of the variational parameters $\alpha_X, \vec{\mu}, K_X$. (In fact, the parameter $\vec{\mu}$ could also be optimised separately for each term $H(A_X|E)$, but this is computationally expensive.)

### J. States saturating the bounds

As discussed in the main part, it is sufficient to only consider the cases $\lambda = 1$(fixed bases) and $\lambda = 1/2$(uniform basis choice) within our protocol. For $\lambda = 1$ the (lower) bound $C^*(S)$ resembles the one from Acìn et. al. [4], which was proven to be optimal in the sense that it can be achieved by an explicit choice of states and measurements.

For $\lambda = 1/2$, we found heuristically that the bound $C^*_{\mathbb{C}^{4\times 4}}(S)$ appears to be saturated by states of the form

$$\rho_a = a\psi^+ + (1-a)\psi^- \quad (58)$$

and measurements given by projectors

$$\Pi_0^{A_0} = \frac{1}{2}\begin{pmatrix} 1+\sin(\tau) & \cos(\tau) \\ \cos(\tau) & 1-\sin(\tau) \end{pmatrix} \qquad \Pi_0^{A_1} = \begin{pmatrix} 1+\sin(\tau) & -\cos(\tau) \\ -\cos(\tau) & 1-\sin(\tau) \end{pmatrix}$$
$$\Pi_0^{B_0} = \frac{1}{2}\begin{pmatrix} 1 & -1 \\ -1 & 1 \end{pmatrix} \qquad \Pi_0^{B_1} = \begin{pmatrix} 1 & 0 \\ 0 & 0 \end{pmatrix}, \tag{59}$$

with parameters $(a, \tau)$ given as optimisers of the program

$$k^* := \min_{a,\tau} \quad 4 - 16(1-a)a\sin(\tau)^2$$
$$s.th.: \quad (4a-2)\cos(\tau) + 2\sin(\tau) = S \quad \frac{1}{2} \leq a \leq 1 \quad 0 \leq \tau \leq \pi/2. \tag{60}$$

Furthermore, we find numerically that the final lower bound on $C(S)$ we obtained appears to be related to $k^*$ via

$$C(S) \geq \log(2) - h_2\left(\frac{1}{2} - \frac{\sqrt{k^*}}{4}\right). \tag{61}$$

However, our heuristic computations of the bound $C^*_{\mathbb{C}^{4\times 4}}(S)$ indicate that it is in fact nonconvex over a large range of values of $S$, and hence the true $C(S)$ bound (for systems of arbitrary dimension) cannot be saturated simply by qubit-qubit systems, but rather only by "block-diagonal combinations" of such systems. In particular, this means that the states described above cannot saturate either the $C(S)$ bound or our final lower bound on it (for all $S$), though it may be possible to do so using "block-diagonal combinations" of them.

## II. SUPPLEMENTARY NOTE 2 - SIMULATION WITH ALL-PHOTONIC SETUP

In this section, we present the simulated asymptotic key rate that one could achieve in an all-photonic experiment. For all-photonic implementations the noise model presented in Ref. [20] gives a more realistic description than the white noise considered in the main text. In this model, the source is modelled by a pair of two-mode squeeze vacua (TMSV) over polarisation modes with mean photon number $\lambda_1$ and $\lambda_2$. Alice and Bob will then set the measurement angles to be $\theta_{A_X}$ and $\theta_{B_Y}$ where $X \in \{0,1\}$ and $Y \in \{0,1,2,3\}$ are their corresponding inputs. Both Alice and Bob will have a pair of single-photon detectors, each with some imperfect quantum efficiency. We let the effective detection efficiency (the total transmittivity between the source and each detector) be $\eta$. Furthermore, we assume that Alice and Bob will give deterministic outputs whenever they observe no detection events. Finally, we also assume that the detectors have negligible dark count rates.

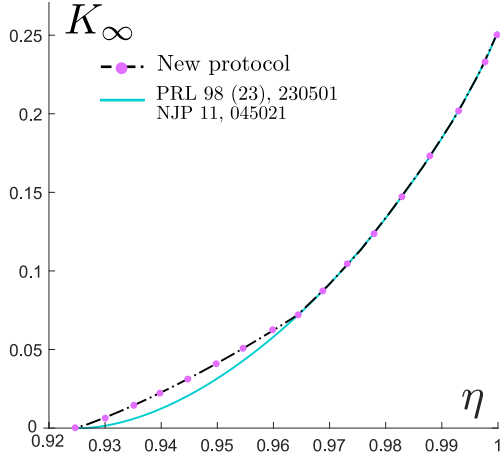

Fig. 6. **Asymptotic key rate for all-photonic experiments.** For this simulation, we use the model presented in Ref. [20] and set the dark count rate of the detectors to zero. We also assumed that Alice and Bob assign deterministic outputs whenever their detectors do not click. We plot the lower bound on the key rate as a function of detection efficiency $\eta$. From the plot, we can see that randomised key basis can improve the achievable key rate in the lower detection efficiency regime.

The corresponding key rate is presented in Fig. 6. To obtain these curves, we optimise the achievable key rates by varying the mean photon number of the TMSV $\lambda_1$ and $\lambda_2$ as well as the measurement angles of Alice and Bob $\theta_{A_X}$ $\theta_{B_Y}$ for each value of $\eta$. Unfortunately, the improvement from employing the randomised key basis protocol is rather limited for the all-photonic experiments. Remarkably, the required detection efficiency to achieve non-zero key rate remains extremely demanding.

## III. SUPPLEMENTARY NOTE 3 - ADDITIONAL LEMMAS AND THEOREMS

**Theorem** 1 DI-Binary Pinsker++: (The following Thm. is a special case of Thm.1 in [10]) Let $Q$ be a (not necessarily rank one) projector on a finite dimensional Hilbert space of dimension $d$ and let $T$ be the pinching channel constructed from $Q$ via

$$T[\rho] = Q\rho Q + (\mathbf{1} - Q)\rho(\mathbf{1} - Q) = \rho + 2Q\rho Q - \{Q, \rho\}. \tag{62}$$

For any state $\rho$, the following inequality holds:

$$D(\rho\|T[\rho]) \geq \log(2) - h_2\left(\frac{1 - \|\rho - T[\rho]\|_1}{2}\right). \tag{63}$$

For the proof of Thm. 1, we will require the following Lemmas:

**Lemma** 1 Gibbs variational principle for operators: Let L be a self-adjoint operator. For any quantum state $\rho$, the following holds:

$$-\operatorname{tr}(\rho L) \geq H(\rho) - \log \operatorname{tr}\left(e^L\right). \tag{64}$$

*Proof.* Consider the thermal state $\sigma = \frac{e^L}{\operatorname{tr}(e^L)}$.

$$-\operatorname{tr}(\rho L) = -\operatorname{tr}\left(\rho \log e^L\right) \tag{65}$$

$$= -\operatorname{tr}\left(\rho\left(\log \frac{e^L}{\operatorname{tr}(e^L)} + \log \operatorname{tr}\left(e^L\right)\right)\right) \tag{66}$$

$$= -\operatorname{tr}\left(\rho \log \frac{e^L}{\operatorname{tr}(e^L)}\right) - \operatorname{tr}\left(\rho \log \operatorname{tr}\left(e^L\right)\right) \tag{67}$$

$$= -\operatorname{tr}(\rho \log \sigma) - \log \operatorname{tr}\left(e^L\right) \tag{68}$$

$$\geq -\operatorname{tr}(\rho \log \rho) - \log \operatorname{tr}\left(e^L\right) \tag{69}$$

$$= H(\rho) - \log \operatorname{tr}\left(e^L\right), \tag{70}$$

where we have used the positivity of the quantum relative entropy $D(\rho\|\sigma) = \operatorname{tr}(\rho \log \rho - \rho \log \sigma)$. $\square$

**Lemma** 2 : For a quantum state $\chi$ and an operator $A$ with spectral radius $r_{\text{spec}}(A) \leq s$, the following relation holds:

$$\operatorname{tr}\left(\chi e^A\right) \leq \cosh(s) + \frac{1}{s}\sinh(s)\operatorname{tr}(\chi A). \tag{71}$$

*Proof.* Firstly, we will find an upper bound to the exponential function in a fixed interval $[-s, s]$. The simplest ansatz is to upper bound it with a linear function, which can be parameterise with $\alpha$ and $\beta$ such that $e^x \leq \alpha + \beta x$. Such an upper bound can be determined by demanding that the points of the exponential function at the boundaries of the interval intersects with the linear function, i.e.

$$\alpha \pm \beta s = e^{\pm s}, \tag{72}$$

which yields $\alpha = \cosh(s)$, $\beta = \frac{1}{s}\sinh(s)$. Hence, we obtain

$$\operatorname{tr}\left(\chi e^A\right) \leq \operatorname{tr}\left(\chi\left[\cosh(s) + \frac{1}{s}\sinh(s)A\right]\right) \tag{73}$$

$$= \cosh(s) + \frac{1}{s}\sinh(s)\operatorname{tr}(\chi A). \tag{74}$$

$\square$

**Lemma** 3 : For any projector $Q$ and quantum state $\rho$, we have

$$\|\rho - T[\rho]\|_1 \leq 1 \tag{75}$$

*Proof.* We use equation (46) and projectors $P$ and $M = 2P - \mathbf{1}$ to rewrite

$$\sup_{\rho Q} \|\rho - T[\rho]\|_1 = 2 \sup_{\rho P Q} \mathrm{tr}\,(P(\rho - T[\rho]))$$
$$= 2 \sup_{\rho P Q} \langle \{P,Q\} - 2QPQ \rangle_\rho$$
$$= 2 \sup_{PQ} \|\{P,Q\} - 2QPQ\|_\infty \qquad (76)$$

Using the usual block decomposition of the algebra generated by two projectors ($P$ and $Q$) we can conclude that the optimal solution of the above will be attained on $2 \times 2$ matrices. We can assume without loss of generality that

$$P = \begin{pmatrix} 1 & 0 \\ 0 & 0 \end{pmatrix} \text{ and } Q = \begin{pmatrix} \cos(\phi)^2 & \cos(\phi)\sin(\phi) \\ \cos(\phi)\sin(\phi) & \sin(\phi)^2 \end{pmatrix}. \qquad (77)$$

Thus, we have

$$\sup_\phi \left\| \left\{ \begin{pmatrix} 1 & 0 \\ 0 & 0 \end{pmatrix}, \begin{pmatrix} \cos(\phi)^2 & \cos(\phi)\sin(\phi) \\ \cos(\phi)\sin(\phi) & \sin(\phi)^2 \end{pmatrix} \right\} \right.$$
$$\left. - 2 \begin{pmatrix} \cos(\phi)^2 & \cos(\phi)\sin(\phi) \\ \cos(\phi)\sin(\phi) & \sin(\phi)^2 \end{pmatrix} \begin{pmatrix} 1 & 0 \\ 0 & 0 \end{pmatrix} \begin{pmatrix} \cos(\phi)^2 & \cos(\phi)\sin(\phi) \\ \cos(\phi)\sin(\phi) & \sin(\phi)^2 \end{pmatrix} \right\|_\infty$$
$$= \frac{1}{2}. \qquad (78)$$

$\square$

*Proof of Pinsker++.* The basic idea of this proof is to find a lower bound on $D(\rho\|T[\rho])$ in terms of $\|\rho - T[\rho]\|_1$ by bounding the expression

$$\inf_\rho D(\rho\|T[\rho]) - \lambda\|\rho - T[\rho]\|_1 \geq c_\lambda \qquad (79)$$

from below by some $c_\lambda$ that only depends on a variational parameter $\lambda \in \mathbb{R}^+$ but not on $\rho$. Given such a $c_\lambda$ for all $\lambda$ will then allow us to obtain a bound on $D$ via an inverse Legendre transformation

$$\forall \rho: \quad D(\rho\|T[\rho]) \geq \sup_{\lambda \in \mathbb{R}^+} c_\lambda + \lambda\|\rho - T[\rho]\|_1 := f(\|\rho - T[\rho]\|_1) \qquad (80)$$

Rewriting (79) with equation (46), $M = 2P - \mathbf{1}$ and Lem. 1 for $L = \log(T[\rho]) + 2\lambda(P - T[P])$ yields the following inequality:

$$D(\rho\|T[\rho]) - \lambda\|\rho - T[\rho]\|_1$$
$$\geq \inf_\rho \inf_{0 \leq P \leq \mathbf{1}} -\mathrm{tr}\,(\rho(\log(T[\rho]) + 2\lambda(P - T[P]))) - H(\rho)$$
$$\geq \inf_\rho \inf_{0 \leq P \leq \mathbf{1}} -\log \mathrm{tr}\,\left(e^{\log(T[\rho]) + 2\lambda(P - T[P])}\right)$$
$$\geq -\log \sup_\rho \sup_{0 \leq P \leq \mathbf{1}} \mathrm{tr}\,\left(e^{\log(T[\rho]) + 2\lambda(P - T[P])}\right)$$
$$\geq -\log \sup_\rho \sup_{0 \leq P \leq \mathbf{1}} \mathrm{tr}\,\left(e^{\log(T[\rho])} e^{2\lambda(P - T[P])}\right)$$
$$= -\log \sup_\rho \sup_{0 \leq P \leq \mathbf{1}} \mathrm{tr}\,\left(T[\rho]\, e^{2\lambda(P - T[P])}\right), \qquad (81)$$

where we have applied the Golden-Thompson inequality $\mathrm{tr}(e^{A+B}) \leq \mathrm{tr}(e^A e^B)$. Next, we use Lem. 2 with $\chi = T[\rho]$, $A = 2\lambda(P - T[P])$, which implies $s = \lambda$ (see Lem. 3). This yields

$$\text{Eq. (81)} \geq -\log\left(\cosh(s) + \sup_\rho \sup_{0 \leq P \leq \mathbf{1}} \frac{2\lambda}{s} \sinh(s)\, \mathrm{tr}\,(T[\rho](P - T[P]))\right)$$
$$= -\log(\cosh(\lambda))$$
$$\qquad (82)$$

16

Hence, we know that

$$D(\rho||T[\rho]) \geq -\log(\cosh(\lambda)) + \lambda\|\rho - T[\rho]\|_1. \tag{83}$$

Since this is true for all $\lambda \geq 0$, the optimal bound can be found by maximizing the right hand side over $\lambda \geq 0$, which gives

$$\lambda_{\max} = \operatorname{arccoth}\left(\frac{1}{\|\rho - T[\rho]\|_1}\right). \tag{84}$$

Inserting it back into (83), where we use the notation $x = \|\rho - T[\rho]\|_1$, yields

$$D(\rho||T[\rho]) \geq -\log\left(\cosh\left(\operatorname{arccoth}\left(\frac{1}{x}\right)\right)\right) + x\operatorname{arccoth}\left(\frac{1}{x}\right) \tag{85}$$

$$= \frac{1}{2}\left(\log(1-x^2) + x\log\left(\frac{1+x}{1-x}\right)\right) \tag{86}$$

$$= \frac{1}{2}\left(\log(1+x) + \log(1-x) + x\log(1+x) - x\log(1-x)\right) \tag{87}$$

$$= \frac{1+x}{2}\log(1+x) + \frac{1-x}{2}\log(1-x) \tag{88}$$

$$= \log(2) + \frac{1+x}{2}\log\left(\frac{1+x}{2}\right) + \frac{1-x}{2}\log\left(\frac{1-x}{2}\right) \tag{89}$$

$$= \log(2) - h_2\left(\frac{1-x}{2}\right), \tag{90}$$

which concludes the proof. □

## IV.  SUPPLEMENTARY NOTE 4 - FINITE-SIZE ANALYSIS

In this section, we give a condensed summary of the finite-size security proof based on the entropy accumulation theorem [21, 22], with the details in an extended writeup [23]. As compared to previous results, the technical contributions from this analysis are as follows. Firstly, it uses tighter finite-size bounds in several points as compared to [24, 25]. Additionally, we perform a different analysis regarding the effects of conditioning on success in the various protocol steps, in order to accommodate practical error-correction protocols that may not have the appropriate form of correctness guarantee required for the analysis in [24]. Furthermore, to apply the entropy accumulation theorem it is necessary to define the protocol steps in such a way that the theorem can be applied to the channels in the protocol — in particular, this requires finding appropriate choices for how to perform the sifting and parameter estimation steps (see below).

### A.  Security definitions

Before beginning the proof, we first need to formalize the security definitions that we aim to achieve. Following [24], we choose the definitions stated below:

**Definition 1.** Consider a DIQKD protocol such that at the end, the honest parties either *accept* (producing keys $K_A$ and $K_B$ of length $\ell_\text{key}$ for Alice and Bob respectively) or *abort* (producing an abort symbol $\perp$ for all parties). It is said to be $\epsilon^\text{com}$-complete and $\epsilon^\text{sou}$-sound if the following properties hold:

- (Completeness) The honest protocol implementation aborts with probability at most $\epsilon^\text{com}$.

- (Soundness) For any implementation of the protocol, we have

$$\Pr[\text{accept}] \frac{1}{2} \left\| \sigma_{K_A K_B E'} - \left( \frac{1}{2^{\ell_\text{key}}} \sum_k |kk\rangle\langle kk|_{K_A K_B} \right) \otimes \sigma_{E'} \right\|_1 \leq \epsilon^\text{sou}, \qquad (91)$$

where $\sigma$ denotes the normalized state conditioned on the protocol accepting, and $E'$ denotes all side-information registers available to the adversary at the end of the protocol.

In the security proof, it is convenient to use the fact that the soundness property is implied by a pair of slightly simpler conditions, as shown in [26]. Specifically, to prove a DIQKD protocol is $\epsilon^\text{sou}$-sound, it suffices to find $\epsilon^\text{cor}_\text{QKD}, \epsilon^\text{sec}_\text{QKD}$ such that $\epsilon^\text{sou} \geq \epsilon^\text{cor}_\text{QKD} + \epsilon^\text{sec}_\text{QKD}$ and the protocol is both $\epsilon^\text{cor}_\text{QKD}$-correct and $\epsilon^\text{sec}_\text{QKD}$-secret, defined as follows:

**Definition 2.** A DIQKD protocol as described above is said to be $\epsilon^\text{cor}_\text{QKD}$-correct and $\epsilon^\text{sec}_\text{QKD}$-secret if the following properties hold:

- (Correctness) For any implementation of the protocol, we have

$$\Pr[K_A \neq K_B \wedge \text{accept}] \leq \epsilon^\text{cor}_\text{QKD}. \qquad (92)$$

- (Secrecy) For any implementation of the protocol, we have

$$\Pr[\text{accept}] \frac{1}{2} \|\sigma_{K_A E'} - \mathbb{U}_{K_A} \otimes \sigma_{E'}\|_1 \leq \epsilon^\text{sec}_\text{QKD}, \qquad (93)$$

where $\sigma$ is as described in Definition 1, and $\mathbb{U}_{K_A}$ denotes the maximally mixed state (i.e. a uniformly random key for Alice).

The parameters in these security definitions are not merely abstract values, but rather they have important operational interpretations. The completeness parameter $\epsilon^\text{com}$ is straightforward — it is simply the probability that the honest devices abort. As for the soundness parameter $\epsilon^\text{sou}$, an important consequence of the definition is regarding the notion of *composability*, as follows [26][2]: suppose one designs a larger protocol (for instance, the one-time-pad protocol) which makes use of an ideal resource that generates a perfectly secret key whenever it does not abort. Furthermore, suppose one proves that when that larger protocol is indeed using the ideal secret-key resource, the probability of some "failure" event

---

[2] Strictly speaking, this interpretation was obtained for the standard QKD setting rather than the DIQKD setting — there are technical issues in formalizing composable security for the latter, because of device-reuse attacks. However, these issues only affect the operational *interpretation* of Definition 1; there are no issues in proving that the protocol satisfies this definition by itself.

(no restrictions need to be imposed on what constitutes a failure, except that it be a well-defined event) is upper-bounded by some value $p_{\text{fail}}$. In that case, it can be shown [26] that if one replaces the ideal secret-key resource with a protocol that is $\epsilon^{\text{sou}}$-sound as defined above, then the probability of the failure event is still upper-bounded by $p_{\text{fail}} + \epsilon^{\text{sou}}$. In other words, the soundness definition allows us to use an $\epsilon^{\text{sou}}$-sound protocol in place of the ideal secret-key resource in a larger protocol, with only an increase of at most $\epsilon^{\text{sou}}$ to the maximum probability of *any* failure event. Further details on this topic can be found in [26].

### B. Security proof sketch

Let $n$ be the total number of rounds in the protocol. Let **A** and **B** denote the output strings that Alice and Bob respectively obtain from the protocol; similarly, let **X** and **Y** denote their input strings. Let **L** denote the syndrome that Alice sends to Bob, and let $E$ denote all the other (quantum) side-information that Eve collects over the course of the protocol (note that it may not have a tensor-product structure since we consider general attacks).

First, we need to describe the error-correction step in a bit more detail. We focus on the case where this step takes the following form: the string **L** consists of two substrings $\mathbf{L}_{\text{EC}}, \mathbf{L}_{\text{h}}$, such that $\mathbf{L}_{\text{EC}}$ is what Bob uses to produce a guess $\tilde{\mathbf{A}}$ for **A**, and $\mathbf{L}_{\text{h}}$ is a 2-universal hash of **A** which Bob uses to check whether his guess is correct (if the hash of his guess $\tilde{\mathbf{A}}$ does not match $\mathbf{L}_{\text{h}}$, Bob announces an abort of the protocol). Furthermore, we shall suppose that before the protocol begins, a constant $\text{EC}_{\text{max}}$ is chosen such that the length of $\mathbf{L}_{\text{EC}}$ is at most $\text{EC}_{\text{max}}$ (we allow the option of using an error-correction procedure that sends a shorter $\mathbf{L}_{\text{EC}}$ if the noise level encountered during execution of the protocol is lower than expected). By using an error-correction step of this form, we shall argue below that the completeness and correctness conditions can be satisfied if we choose suitable values for $\text{EC}_{\text{max}}$ and the length of $\mathbf{L}_{\text{h}}$.

Also, for the parameter-estimation analysis, instead of using the CHSH value in terms of correlators, it is easier to consider the winning probability of the *CHSH game*. Specifically, view Alice and Bob's outputs as taking values in $\{0,1\}$ rather than $\{+1,-1\}$, and for outputs $a,b \in \{0,1\}$ from inputs $x \in \{0,1\}$, $y \in \{2,3\}$ (in a test round), we say that Alice and Bob win the CHSH game if $a \oplus b = x \cdot (y-2)$. The probability $w$ of winning the CHSH game (given uniformly random inputs) is related to the correlator-based CHSH value $S$ by the simple conversion $S = 8w - 4$. We shall suppose that the honest devices are IID and achieve some CHSH winning probability $w_{\text{exp}}$ in each individual round.

For simplicity, we shall focus on the case where the protocol is performed with $p = 1/2$, since our results indicate that this is optimal at higher noise levels. In that case, the inputs in the test rounds are uniformly random, so Alice and Bob can indeed be considered to be playing the CHSH game in those rounds. Given this, it is easier to instead consider a slightly different version of the parameter-estimation step, which differs somewhat from more commonly studied QKD protocols, but facilitates a technical application of the entropy accumulation theorem to prove security against general attacks [23]. Specifically, parameter estimation is performed as follows: introducing a tolerance value $\delta_{\text{tol}}$, Bob will check that the number of test rounds that won the CHSH game is at least $(w_{\text{exp}} - \delta_{\text{tol}})qn$, and also that the number of test rounds that lost the CHSH game is at most $(1 - w_{\text{exp}} + \delta_{\text{tol}})qn$. (Performing both of these checks allows for a technical optimization in the finite-size security analysis; see [23]. Also, note that we do not "normalize" by dividing by the number of test rounds that actually occurred — the $qn$ factor is computed directly from the protocol parameters. Our results indicate [23] that this form of parameter estimation incurs certain losses compared to the more standard form, but it remains an open question whether the latter can be formalised in a manner compatible with the entropy accumulation theorem.) If either of these conditions is not satisfied, Bob aborts the protocol at that step.

In the previous sections, we have derived a lower bound on $\lambda H(A_0|E) + (1-\lambda) H(A_1|E)$ as a function of the CHSH value $S$ (for the choice $p = 1/2$ we have $\lambda = 1 - \lambda = 1/2$). Now, let $r(w)$ denote such a lower bound, but in terms of the CHSH winning probability $w$ instead, and with the additional condition that it be an affine function (this could be obtained in principle as part of the Legendre-transform process we described in Sec. IG, though in practice we found it more efficient to use the approach described in [23]). With this in mind, we now define a function

$$g(w) = \frac{1+q}{2} r(w), \qquad (94)$$

which can be informally interpreted as a lower bound on the entropy "accumulated" in one round of the protocol. (The prefactor is to account for the fact that Alice and Bob sift out generation rounds which they have chosen different inputs[3]. To maintain compatibility with entropy accumulation, we interpret the sifting step as Alice and Bob setting their outcomes to some deterministic values whenever their inputs differ, rather than truly "discarding" those rounds.)

With this in mind, we can state the security guarantees of the protocol [23]:

---

[3] To optimize the keyrates, we include the test rounds in the contribution to the entropy; see [23].

**Theorem 2 :** Take any $\epsilon_{\text{EC}}^{\text{com}}, \epsilon_{\text{PE}}^{\text{com}}, \epsilon_{\text{EA}}, \epsilon_{\text{PA}}, \epsilon_{\text{h}}, \epsilon_s, \epsilon'_s, \epsilon''_s \in (0,1]$ such that $\epsilon_s > \epsilon'_s + 2\epsilon''_s$, and any $\alpha \in (1,2)$, $\alpha' \in (1, 1+2/V')$, $\beta \in [g(0), g(1)]$, $q \in (0,1)$, where $V' = 2\log 7$. For $p = 1/2$, the described protocol is $(\epsilon_{\text{EC}}^{\text{com}} + \epsilon_{\text{PE}}^{\text{com}})$-complete and $(\max\{\epsilon_{\text{EA}}, \epsilon_{\text{PA}} + 2\epsilon_s\} + 2\epsilon_{\text{h}})$-sound when performed with $\text{EC}_{\max}$ satisfying Eq. (99), and $\delta_{\text{tol}}, \ell_{\text{key}}$ satisfying

$$\epsilon_{\text{PE}}^{\text{com}} \geq B_{n, q w_{\exp}}(\lfloor (w_{\exp} - \delta_{\text{tol}}) q n \rfloor) + B_{n, 1-q+qw_{\exp}}(\lfloor (1 - q + w_{\exp} q - \delta_{\text{tol}} q) n \rfloor), \tag{95}$$

$$\ell_{\text{key}} \leq ng(w_{\exp} - \delta_{\text{tol}}) - n\frac{(\alpha-1)\ln 2}{2}V^2 - n(\alpha-1)^2 K_\alpha^2 - nq - n\left(\frac{\alpha'-1}{4}\right)V'^2$$

$$- \frac{\vartheta_{\epsilon'_s}}{\alpha-1} - \frac{\vartheta_{\epsilon''_s}}{\alpha'-1} - \left(\frac{\alpha}{\alpha-1} + \frac{\alpha'}{\alpha'-1} - 2\right)\log\frac{1}{\epsilon_{\text{EA}}} - 3\vartheta_{\epsilon_s - \epsilon'_s - 2\epsilon''_s}$$

$$- \text{EC}_{\max} - \left\lceil \log\left(\frac{1}{\epsilon_{\text{h}}}\right)\right\rceil - 2\log\frac{1}{\epsilon_{\text{PA}}} + 2, \tag{96}$$

where $V$ and $K_\alpha$ are constants (depending on $q, g, \beta$ and $\alpha$) that can be explicitly computed [22, 23], $\vartheta_\epsilon$ is defined as

$$\vartheta_\epsilon = \log\frac{1}{1 - \sqrt{1 - \epsilon^2}}, \tag{97}$$

and $B_{n,p}(k)$ denotes the cumulative distribution function of a binomial distribution with parameters $(n, p)$:

$$B_{n,p}(k) = \Pr_{X \sim \text{Binom}(n,p)}[X \leq k]. \tag{98}$$

The variables listed at the start of Theorem 2 can be considered to be variational parameters that should be chosen to optimize the keyrate as much as possible. To get a sense of the asymptotic scaling, it can be shown [21–23] that as $n \to \infty$, we can choose these parameters in such a manner that $\delta_{\text{tol}}, q \to 0$ and the key-length expression (96) is dominated by the terms $ng(w_{\exp} - \delta_{\text{tol}})$ and $\text{EC}_{\max}$. By taking Eq. (99) for the latter, this yields an asymptotic keyrate of essentially[4] $(1/2)\left(\sum_x H(A_x|E) - \sum_x H(A_x|B_x)\right)$ as expected.

We now sketch the proof of the above theorem, by considering those variational parameters to take some fixed values, and then showing that indeed the desired security claims (as functions of these values) hold.

**Completeness:** To show that this condition holds, we impose the requirement that $\text{EC}_{\max}$ is long enough such that for the honest devices, Bob can use $\mathbf{L}_{\text{EC}}$ and $\mathbf{B}$ to produce a guess for $\mathbf{A}$ that is correct with probability at least $1 - \epsilon_{\text{EC}}^{\text{com}}$. From the results of [21, 27], this is possible as long as $\text{EC}_{\max}$ is chosen such that[5]

$$\text{EC}_{\max} \geq nh_{\text{hon}} + \sqrt{n}\,(2\log 5)\sqrt{\log\frac{2}{\tilde{\epsilon}_s^2}} + 2\log\frac{1}{\epsilon_{\text{EC}}^{\text{com}} - \tilde{\epsilon}_s} + 4, \tag{99}$$

where $\tilde{\epsilon}_s \in [0, \epsilon_{\text{EC}}^{\text{com}})$ is another parameter that can be optimized over, and

$$h_{\text{hon}} = \frac{1-q}{4}\left(H(A_0|B_0)_{\text{hon}} + H(A_1|B_1)_{\text{hon}}\right) + q h_2(w_{\exp}), \tag{100}$$

where the subscript $_{\text{hon}}$ denotes that the terms should be computed with respect to the honest behaviour, and for simplicity we have assumed that the honest devices win the CHSH game with the same probability $w_{\exp}$ for all input pairs. Having imposed this requirement (that Bob's guess is correct with probability at least $1 - \epsilon_{\text{EC}}^{\text{com}}$) on the honest protocol, we see that the probability that it aborts during the error-correction step (because the hashes of $\tilde{\mathbf{A}}$ and $\mathbf{A}$ do not match) is at most $\epsilon_{\text{EC}}^{\text{com}}$. Furthermore, since the honest devices win the CHSH game with probability $w_{\exp}$, and the parameter-estimation step accepts when the number of test rounds that won (resp. lost) the CHSH game is at least $(w_{\exp} - \delta_{\text{tol}})qn$ (resp. at most $(1 - w_{\exp} + \delta_{\text{tol}})qn$), it is not hard to show that the probability that the honest protocol aborts during the parameter-estimation step is at most $\epsilon_{\text{PE}}^{\text{com}}$ as defined in Eq. (95) (because in the honest protocol, the number of test rounds that win/lose the CHSH game follows a binomial distribution). Applying the union bound to these different ways the protocol could abort, we see the protocol is $(\epsilon_{\text{EC}}^{\text{com}} + \epsilon_{\text{PE}}^{\text{com}})$-complete as claimed.

---

[4] Technically, $H(A_x|E)$ and $H(A_x|B_x)$ refer to single-round quantities and are hence only well-defined when the rounds are IID, i.e. the values of these quantities are the same in all rounds. However, the asymptotic rate we have described for Theorem 2 is essentially analogous in the sense that $(1/2)\sum_x H(A_x|E)$ is replaced by a lower bound $g(w_{\exp} - \delta_{\text{tol}})$ that, informally speaking, holds for all rounds. (As for the $H(A_x|B_x)$ terms, they are computed with respect to the honest behaviour, which is indeed IID.)

[5] Other error-correction procedures are possible in principle; see [23] for further discussion.

**Soundness:** To prove the claim of $(\max\{\epsilon_{\text{EA}}, \epsilon_{\text{PA}} + 2\epsilon_s\} + 2\epsilon_{\text{h}})$-soundness, we argue the protocol is $\epsilon_{\text{h}}$-correct and $(\max\{\epsilon_{\text{EA}}, \epsilon_{\text{PA}} + 2\epsilon_s\} + \epsilon_{\text{h}})$-secret, as follows:

**Correctness:** To satisfy this condition, we shall set the length of $\mathbf{L}_{\text{h}}$ to be $\lceil \log(1/\epsilon_{\text{h}}) \rceil$. By the defining property of 2-universal hashing, for a hash of this length, the probability that two different strings hash to the same value is at most $\epsilon_{\text{h}}$. Since $K_A$ and $K_B$ are produced from $\mathbf{A}$ and $\tilde{\mathbf{A}}$ respectively, a straightforward calculation shows that this implies $\Pr[K_A \neq K_B \wedge \text{accept}] \leq \epsilon_{\text{h}}$ as well, and hence we have that the protocol is $\epsilon_{\text{h}}$-correct.

**Secrecy:** The main challenge is in proving this condition holds. To slightly improve the keyrates, in proving Theorem 2 we impose (see [23, 24]) that Alice's test-round outputs are not sent in plaintext to Bob, but rather they are compressed into an error-correction syndrome as well. In that case, strictly speaking Bob is performing parameter estimation using his guess $\tilde{\mathbf{A}}$ rather than $\mathbf{A}$ itself, but it can be shown [23] that roughly speaking, the fact that we have $\epsilon_{\text{h}}$-correctness as shown above implies that we can perform the analysis conditioned on $\tilde{\mathbf{A}} = \mathbf{A}$, at the cost of increasing the secrecy parameter by $\epsilon_{\text{h}}$. We will not further discuss this point here, except to remember to add $\epsilon_{\text{h}}$ to the secrecy parameter at the end.

At the end of the protocol Eve holds the registers $\mathbf{XYL}E$, and the final privacy amplification step in the protocol was performed by Alice on the register $\mathbf{A}$ (and by Bob on $\tilde{\mathbf{A}}$). The fundamental property of privacy amplification is basically that it produces a final key satisfying the secrecy definition, with length related to the smoothed min-entropy of $\mathbf{A}$ conditioned on $\mathbf{XYL}E$. More specifically [23, 28], as long as we set the length of the privacy-amplification output (i.e. the final key) to satisfy

$$\ell_{\text{key}} \leq H_{\min}^{\epsilon_s}(\mathbf{A}|\mathbf{XYL}E) + 2 - 2\log\frac{1}{\epsilon_{\text{PA}}}, \tag{101}$$

we will get an $(\epsilon_{\text{PA}} + 2\epsilon_s)$-secret key (putting aside several technicalities about conditioning on the various steps accepting). Hence to find a bound on the final achievable key length, it suffices to give a lower bound on $H_{\min}^{\epsilon_s}(\mathbf{A}|\mathbf{XYL}E)$. To do so, we first apply a chain rule [29] to get

$$H_{\min}^{\epsilon_s}(\mathbf{A}|\mathbf{XYL}E) \geq H_{\min}^{\epsilon_s}(\mathbf{A}|\mathbf{XY}E) - \text{len}(\mathbf{L})$$
$$\geq H_{\min}^{\epsilon_s}(\mathbf{A}|\mathbf{XY}E) - \text{EC}_{\max} - \left\lceil \log\left(\frac{1}{\epsilon_{\text{h}}}\right) \right\rceil, \tag{102}$$

reducing our task to bounding $H_{\min}^{\epsilon_s}(\mathbf{A}|\mathbf{XY}E)$. This is achieved by using the entropy accumulation theorem, which yields a lower bound on the smoothed min-entropy of a state produced by a sequence of channels (which in this case we basically take to correspond to the rounds of the protocol). Specifically, the entropy accumulation theorem can be used to show that one of the following must be true: (1) the probability of the parameter-estimation step accepting is at most $\epsilon_{\text{EA}}$, or (2) conditioned on accepting, we have the lower bound

$$H_{\min}^{\epsilon_s}(\mathbf{A}|\mathbf{XY}E) > ng(w_{\exp} - \delta_{\text{tol}}) - n\frac{(\alpha-1)\ln 2}{2}V^2 - n(\alpha-1)^2 K_\alpha^2 - nq - n\left(\frac{\alpha'-1}{4}\right)V'^2$$
$$- \frac{\vartheta_{\epsilon'_s}}{\alpha-1} - \frac{\vartheta_{\epsilon''_s}}{\alpha'-1} - \left(\frac{\alpha}{\alpha-1} + \frac{\alpha'}{\alpha'-1}\right)\log\frac{1}{\epsilon_{\text{EA}}} - 3\vartheta_{\epsilon_s - \epsilon'_s - 2\epsilon''_s}. \tag{103}$$

In case (1), the secrecy condition immediately holds with secrecy parameter $\epsilon_{\text{EA}}$. In case (2), the sequence of equations (101)–(103) tells us that we can get an $(\epsilon_{\text{PA}} + 2\epsilon_s)$-secret key with length given by (96)[6]. Recalling to add the $\epsilon_{\text{h}}$ correction because parameter estimation is performed with $\tilde{\mathbf{A}}$ rather than $\mathbf{A}$, we conclude the protocol is $(\max\{\epsilon_{\text{EA}}, \epsilon_{\text{PA}} + 2\epsilon_s\} + \epsilon_{\text{h}})$-secret.

In the above, we have described the key rates for general attacks. If we make the assumption of collective attacks, then the completeness and correctness proofs proceed in exactly the same way — only the secrecy proof needs to be modified. The main modification (ignoring some technical details) is that the bound (103) is replaced with one based on the *asymptotic equipartition property* (we use the version presented in [21]), which holds for IID states. The resulting rates are higher than those resulting from (103), since the asymptotic equipartition property exploits the assumption of

---

[6] To be precise, the expression (96) has a slightly improved dependence on $\epsilon_{\text{EA}}$ by keeping track of the event conditioning in a more precise statement of privacy amplification (Eq. (101)), but we defer the details to [23].


\end{widetext}

\begin{thebibliography}{10}
\expandafter\ifx\csname url\endcsname\relax
  \def\url#1{\texttt{#1}}\fi
\expandafter\ifx\csname urlprefix\endcsname\relax\def\urlprefix{URL }\fi
\providecommand{\bibinfo}[2]{#2}
\providecommand{\eprint}[2][]{\url{#2}}

\bibitem{mayers1998quantum}
\bibinfo{author}{Mayers, D.} \& \bibinfo{author}{Yao, A.}
\newblock \bibinfo{title}{Quantum cryptography with imperfect apparatus}.
\newblock In \emph{\bibinfo{booktitle}{Proceedings 39th Annual Symposium on
  Foundations of Computer Science (Cat. No. 98CB36280)}},
  \bibinfo{pages}{503--509} (\bibinfo{organization}{IEEE},
  \bibinfo{year}{1998}).

\bibitem{pironio}
\bibinfo{author}{Pironio, S.} \emph{et~al.}
\newblock \bibinfo{title}{Device-independent quantum key distribution secure
  against collective attacks}.
\newblock \emph{\bibinfo{journal}{New Journal of Physics}}
  \textbf{\bibinfo{volume}{11}}, \bibinfo{pages}{045021}
  (\bibinfo{year}{2009}).

\bibitem{acin2007device}
\bibinfo{author}{Ac{\'\i}n, A.} \emph{et~al.}
\newblock \bibinfo{title}{Device-independent security of quantum cryptography
  against collective attacks}.
\newblock \emph{\bibinfo{journal}{Physical Review Letters}}
  \textbf{\bibinfo{volume}{98}}, \bibinfo{pages}{230501}
  (\bibinfo{year}{2007}).

\bibitem{barrett2005no}
\bibinfo{author}{Barrett, J.}, \bibinfo{author}{Hardy, L.} \&
  \bibinfo{author}{Kent, A.}
\newblock \bibinfo{title}{No signaling and quantum key distribution}.
\newblock \emph{\bibinfo{journal}{Physical Review Letters}}
  \textbf{\bibinfo{volume}{95}}, \bibinfo{pages}{010503}
  (\bibinfo{year}{2005}).

\bibitem{reichardt2013classical}
\bibinfo{author}{Reichardt, B.~W.}, \bibinfo{author}{Unger, F.} \&
  \bibinfo{author}{Vazirani, U.}
\newblock \bibinfo{title}{Classical command of quantum systems}.
\newblock \emph{\bibinfo{journal}{Nature}} \textbf{\bibinfo{volume}{496}},
  \bibinfo{pages}{456--460} (\bibinfo{year}{2013}).

\bibitem{clauser1969proposed}
\bibinfo{author}{Clauser, J.~F.}, \bibinfo{author}{Horne, M.~A.},
  \bibinfo{author}{Shimony, A.} \& \bibinfo{author}{Holt, R.~A.}
\newblock \bibinfo{title}{Proposed experiment to test local hidden-variable
  theories}.
\newblock \emph{\bibinfo{journal}{Physical Review Letters}}
  \textbf{\bibinfo{volume}{23}}, \bibinfo{pages}{880} (\bibinfo{year}{1969}).

\bibitem{Bell1964}
\bibinfo{author}{Bell, J.~S.}
\newblock \bibinfo{title}{On the {E}instein-{P}odolsky-{R}osen paradox}.
\newblock \emph{\bibinfo{journal}{Physics}} \textbf{\bibinfo{volume}{1}},
  \bibinfo{pages}{195--200} (\bibinfo{year}{1964}).

\bibitem{miller2016robust}
\bibinfo{author}{Miller, C.~A.} \& \bibinfo{author}{Shi, Y.}
\newblock \bibinfo{title}{Robust protocols for securely expanding randomness
  and distributing keys using untrusted quantum devices}.
\newblock \emph{\bibinfo{journal}{Journal of the ACM}}
  \textbf{\bibinfo{volume}{63}}, \bibinfo{pages}{1--63} (\bibinfo{year}{2016}).

\bibitem{vazirani2019fully}
\bibinfo{author}{Vazirani, U.} \& \bibinfo{author}{Vidick, T.}
\newblock \bibinfo{title}{Fully device independent quantum key distribution}.
\newblock \emph{\bibinfo{journal}{Communications of the ACM}}
  \textbf{\bibinfo{volume}{62}}, \bibinfo{pages}{133--133}
  (\bibinfo{year}{2019}).

\bibitem{arnon2019simple}
\bibinfo{author}{Arnon-Friedman, R.}, \bibinfo{author}{Renner, R.} \&
  \bibinfo{author}{Vidick, T.}
\newblock \bibinfo{title}{Simple and tight device-independent security proofs}.
\newblock \emph{\bibinfo{journal}{SIAM Journal on Computing}}
  \textbf{\bibinfo{volume}{48}}, \bibinfo{pages}{181--225}
  (\bibinfo{year}{2019}).

\bibitem{PR21}
\bibinfo{author}{Portmann, C.} \& \bibinfo{author}{Renner, R.}
\newblock \bibinfo{title}{{Security in Quantum Cryptography}}.
\newblock \emph{\bibinfo{journal}{arXiv:2102.00021 [quant-ph]}}
  (\bibinfo{year}{2021}).
\newblock \urlprefix\url{https://arxiv.org/abs/2102.00021v1}.

\bibitem{barrett2013memory}
\bibinfo{author}{Barrett, J.}, \bibinfo{author}{Colbeck, R.} \&
  \bibinfo{author}{Kent, A.}
\newblock \bibinfo{title}{Memory attacks on device-independent quantum
  cryptography}.
\newblock \emph{\bibinfo{journal}{Physical Review Letters}}
  \textbf{\bibinfo{volume}{110}}, \bibinfo{pages}{010503}
  (\bibinfo{year}{2013}).

\bibitem{vazirani2014robust}
\bibinfo{author}{Vazirani, U.} \& \bibinfo{author}{Vidick, T.}
\newblock \bibinfo{title}{Robust device independent quantum key distribution}.
\newblock In \emph{\bibinfo{booktitle}{Proceedings of the 5th conference on
  Innovations in theoretical computer science}}, \bibinfo{pages}{35--36}
  (\bibinfo{year}{2014}).

\bibitem{Hensen2015}
\bibinfo{author}{Hensen, B.} \emph{et~al.}
\newblock \bibinfo{title}{{Loophole-free Bell inequality violation using
  electron spins separated by 1.3 kilometres}}.
\newblock \emph{\bibinfo{journal}{Nature}} \textbf{\bibinfo{volume}{526}},
  \bibinfo{pages}{682--686} (\bibinfo{year}{2015}).

\bibitem{Giustina2015}
\bibinfo{author}{Giustina, M.} \emph{et~al.}
\newblock \bibinfo{title}{{Significant-Loophole-Free Test of Bell's Theorem
  with Entangled Photons}}.
\newblock \emph{\bibinfo{journal}{Physical Review Letters}}
  \textbf{\bibinfo{volume}{115}}, \bibinfo{pages}{250401}
  (\bibinfo{year}{2015}).

\bibitem{Shalm2015}
\bibinfo{author}{Shalm, L.~K.} \emph{et~al.}
\newblock \bibinfo{title}{Strong {Loophole}-{Free} {Test} of {Local}
  {Realism}}.
\newblock \emph{\bibinfo{journal}{Physical Review Letters}}
  \textbf{\bibinfo{volume}{115}}, \bibinfo{pages}{250402}
  (\bibinfo{year}{2015}).

\bibitem{Rosenfeld2017}
\bibinfo{author}{Rosenfeld, W.} \emph{et~al.}
\newblock \bibinfo{title}{{Event-ready Bell test using entangled atoms
  simultaneously closing detection and locality loopholes}}.
\newblock \emph{\bibinfo{journal}{Physical Review Letters}}
  \textbf{\bibinfo{volume}{119}}, \bibinfo{pages}{010402}
  (\bibinfo{year}{2017}).

\bibitem{Murta2018}
\bibinfo{author}{Murta, G.}, \bibinfo{author}{van Dam, S.~B.},
  \bibinfo{author}{Ribeiro, J.}, \bibinfo{author}{Hanson, R.} \&
  \bibinfo{author}{Wehner, S.}
\newblock \bibinfo{title}{Towards a realization of device-independent quantum
  key distribution}.
\newblock \emph{\bibinfo{journal}{Quantum Science and Technology}}
  \textbf{\bibinfo{volume}{4}}, \bibinfo{pages}{035011} (\bibinfo{year}{2019}).

\bibitem{gisin2010proposal}
\bibinfo{author}{Gisin, N.}, \bibinfo{author}{Pironio, S.} \&
  \bibinfo{author}{Sangouard, N.}
\newblock \bibinfo{title}{Proposal for implementing device-independent quantum
  key distribution based on a heralded qubit amplifier}.
\newblock \emph{\bibinfo{journal}{Physical Review Letters}}
  \textbf{\bibinfo{volume}{105}}, \bibinfo{pages}{070501}
  (\bibinfo{year}{2010}).

\bibitem{Moroder2011}
\bibinfo{author}{Curty, M.} \& \bibinfo{author}{Moroder, T.}
\newblock \bibinfo{title}{Heralded-qubit amplifiers for practical
  device-independent quantum key distribution}.
\newblock \emph{\bibinfo{journal}{Phys. Rev. A}} \textbf{\bibinfo{volume}{84}},
  \bibinfo{pages}{010304} (\bibinfo{year}{2011}).

\bibitem{Kolodynski2020}
\bibinfo{author}{Ko{\l}ody{\'{n}}ski, J.} \emph{et~al.}
\newblock \bibinfo{title}{Device-independent quantum key distribution with
  single-photon sources}.
\newblock \emph{\bibinfo{journal}{{Quantum}}} \textbf{\bibinfo{volume}{4}},
  \bibinfo{pages}{260} (\bibinfo{year}{2020}).
\newblock \urlprefix\url{https://doi.org/10.22331/q-2020-04-30-260}.

\bibitem{ac1}
\bibinfo{author}{Cabello, A.} \& \bibinfo{author}{Sciarrino, F.}
\newblock \bibinfo{title}{Loophole-free bell test based on local
  precertification of photon’s presence}.
\newblock \emph{\bibinfo{journal}{Physical Review X}}
  \textbf{\bibinfo{volume}{2}}, \bibinfo{pages}{021010} (\bibinfo{year}{2012}).

\bibitem{ac2}
\bibinfo{author}{Meyer-Scott, E.} \emph{et~al.}
\newblock \bibinfo{title}{Certifying the presence of a photonic qubit by
  splitting it in two}.
\newblock \emph{\bibinfo{journal}{Physical Review Letters}}
  \textbf{\bibinfo{volume}{116}}, \bibinfo{pages}{070501}
  (\bibinfo{year}{2016}).

\bibitem{limdevice}
\bibinfo{author}{Lim, C. C.-W.}, \bibinfo{author}{Portmann, C.},
  \bibinfo{author}{Tomamichel, M.}, \bibinfo{author}{Renner, R.} \&
  \bibinfo{author}{Gisin, N.}
\newblock \bibinfo{title}{Device-independent quantum key distribution with
  local bell test}.
\newblock \emph{\bibinfo{journal}{Physical Review X}}
  \textbf{\bibinfo{volume}{3}}, \bibinfo{pages}{031006} (\bibinfo{year}{2013}).

\bibitem{tan2020advantage}
\bibinfo{author}{Tan, E. Y.-Z.}, \bibinfo{author}{Lim, C. C.-W.} \&
  \bibinfo{author}{Renner, R.}
\newblock \bibinfo{title}{Advantage distillation for device-independent quantum
  key distribution}.
\newblock \emph{\bibinfo{journal}{Physical Review Letters}}
  \textbf{\bibinfo{volume}{124}}, \bibinfo{pages}{020502}
  (\bibinfo{year}{2020}).

\bibitem{scarani2009security}
\bibinfo{author}{Scarani, V.} \emph{et~al.}
\newblock \bibinfo{title}{The security of practical quantum key distribution}.
\newblock \emph{\bibinfo{journal}{Reviews of Modern Physics}}
  \textbf{\bibinfo{volume}{81}}, \bibinfo{pages}{1301} (\bibinfo{year}{2009}).

\bibitem{hall11}
\bibinfo{author}{Hall, M. J.~W.}
\newblock \bibinfo{title}{{Relaxed Bell inequalities and Kochen-Specker
  theorems}}.
\newblock \emph{\bibinfo{journal}{Phys. Rev. A}} \textbf{\bibinfo{volume}{84}},
  \bibinfo{pages}{022102} (\bibinfo{year}{2011}).
\newblock \urlprefix\url{http://link.aps.org/doi/10.1103/PhysRevA.84.022102}.

\bibitem{Fine82}
\bibinfo{author}{Fine, A.}
\newblock \bibinfo{title}{Joint distributions, quantum correlations, and
  commuting observables}.
\newblock \emph{\bibinfo{journal}{Journal of Mathematical Physics}}
  \textbf{\bibinfo{volume}{23}}, \bibinfo{pages}{1306--1310}
  (\bibinfo{year}{1982}).

\bibitem{busch}
\bibinfo{author}{Busch, P.}
\newblock \bibinfo{title}{Indeterminacy relations and simultaneous measurements
  in quantum theory}.
\newblock \emph{\bibinfo{journal}{International Journal of Theoretical
  Physics}} \textbf{\bibinfo{volume}{24}}, \bibinfo{pages}{63--92}
  (\bibinfo{year}{1985}).

\bibitem{Wolf2009}
\bibinfo{author}{Wolf, M.~M.}, \bibinfo{author}{Perez-Garcia, D.} \&
  \bibinfo{author}{Fernandez, C.}
\newblock \bibinfo{title}{Measurements incompatible in quantum theory cannot be
  measured jointly in any other no-signaling theory}.
\newblock \emph{\bibinfo{journal}{Phys. Rev. Lett.}}
  \textbf{\bibinfo{volume}{103}}, \bibinfo{pages}{230402}
  (\bibinfo{year}{2009}).
\newblock
  \urlprefix\url{https://link.aps.org/doi/10.1103/PhysRevLett.103.230402}.

\bibitem{HorodeckiRMP}
\bibinfo{author}{Horodecki, R.}, \bibinfo{author}{Horodecki, P.},
  \bibinfo{author}{Horodecki, M.} \& \bibinfo{author}{Horodecki, K.}
\newblock \bibinfo{title}{Quantum entanglement}.
\newblock \emph{\bibinfo{journal}{Rev. Mod. Phys.}}
  \textbf{\bibinfo{volume}{81}}, \bibinfo{pages}{865--942}
  (\bibinfo{year}{2009}).
\newblock \urlprefix\url{https://link.aps.org/doi/10.1103/RevModPhys.81.865}.

\bibitem{Lo2005}
\bibinfo{author}{Lo, H.-K.}, \bibinfo{author}{Chau, H.~F.} \&
  \bibinfo{author}{Ardehali, M.}
\newblock \bibinfo{title}{Efficient quantum key distribution scheme and a proof
  of its unconditional security}.
\newblock \emph{\bibinfo{journal}{Journal of Cryptology}}
  \textbf{\bibinfo{volume}{18}}, \bibinfo{pages}{133--165}
  (\bibinfo{year}{2005}).
\newblock \urlprefix\url{https://doi.org/10.1007/s00145-004-0142-y}.

\bibitem{berta}
\bibinfo{author}{Berta, M.}, \bibinfo{author}{Christandl, M.},
  \bibinfo{author}{Colbeck, R.}, \bibinfo{author}{Renes, J.~M.} \&
  \bibinfo{author}{Renner, R.}
\newblock \bibinfo{title}{The uncertainty principle in the presence of quantum
  memory}.
\newblock \emph{\bibinfo{journal}{Nature Physics}}
  \textbf{\bibinfo{volume}{6}}, \bibinfo{pages}{659} (\bibinfo{year}{2010}).

\bibitem{eureview2}
\bibinfo{author}{Coles, P.}, \bibinfo{author}{Berta, M.},
  \bibinfo{author}{Tomamichel, M.} \& \bibinfo{author}{Wehner, S.}
\newblock \bibinfo{title}{Entropic uncertainty relations and their
  applications}.
\newblock \emph{\bibinfo{journal}{Reviews of Modern Physics}}
  \textbf{\bibinfo{volume}{89}} (\bibinfo{year}{2017}).
\newblock \bibinfo{note}{And
  \href{https://arxiv.org/abs/1511.04857}{arXiv:1511.04857}}.

\bibitem{henkel2010highly}
\bibinfo{author}{Henkel, F.} \emph{et~al.}
\newblock \bibinfo{title}{Highly efficient state-selective submicrosecond
  photoionization detection of single atoms}.
\newblock \emph{\bibinfo{journal}{Physical Review Letters}}
  \textbf{\bibinfo{volume}{105}}, \bibinfo{pages}{253001}
  (\bibinfo{year}{2010}).

\bibitem{matsukevich2008bell}
\bibinfo{author}{Matsukevich, D.~N.}, \bibinfo{author}{Maunz, P.},
  \bibinfo{author}{Moehring, D.~L.}, \bibinfo{author}{Olmschenk, S.} \&
  \bibinfo{author}{Monroe, C.}
\newblock \bibinfo{title}{{Bell Inequality Violation with Two Remote Atomic
  Qubits}}.
\newblock \emph{\bibinfo{journal}{Physical Review Letters}}
  \textbf{\bibinfo{volume}{100}}, \bibinfo{pages}{150404}
  (\bibinfo{year}{2008}).

\bibitem{pironio2010random}
\bibinfo{author}{Pironio, S.} \emph{et~al.}
\newblock \bibinfo{title}{{Random numbers certified by Bell’s theorem}}.
\newblock \emph{\bibinfo{journal}{Nature}} \textbf{\bibinfo{volume}{464}},
  \bibinfo{pages}{1021--1024} (\bibinfo{year}{2010}).

\bibitem{giustina2013bell}
\bibinfo{author}{Giustina, M.} \emph{et~al.}
\newblock \bibinfo{title}{Bell violation using entangled photons without the
  fair-sampling assumption}.
\newblock \emph{\bibinfo{journal}{Nature}} \textbf{\bibinfo{volume}{497}},
  \bibinfo{pages}{227--230} (\bibinfo{year}{2013}).

\bibitem{christensen2013detection}
\bibinfo{author}{Christensen, B.} \emph{et~al.}
\newblock \bibinfo{title}{{Detection-Loophole-Free Test of Quantum Nonlocality,
  and Applications}}.
\newblock \emph{\bibinfo{journal}{Physical Review Letters}}
  \textbf{\bibinfo{volume}{111}}, \bibinfo{pages}{130406}
  (\bibinfo{year}{2013}).

\bibitem{Liu18}
\bibinfo{author}{Liu, Y.} \emph{et~al.}
\newblock \bibinfo{title}{Device-independent quantum random-number generation}.
\newblock \emph{\bibinfo{journal}{Nature}} \textbf{\bibinfo{volume}{562}},
  \bibinfo{pages}{548--551} (\bibinfo{year}{2018}).

\bibitem{Liu19}
\bibinfo{author}{Liu, W.-Z.} \emph{et~al.}
\newblock \bibinfo{title}{Device-independent randomness expansion against
  quantum side information}.
\newblock \emph{\bibinfo{journal}{arXiv:1912.11159}}  (\bibinfo{year}{2019}).

\bibitem{Li19}
\bibinfo{author}{Li, M.-H.} \emph{et~al.}
\newblock \bibinfo{title}{Experimental realization of device-independent
  quantum randomness expansion}.
\newblock \emph{\bibinfo{journal}{arXiv:1902.07529}}  (\bibinfo{year}{2019}).

\bibitem{zukowski1993event}
\bibinfo{author}{{\.Z}ukowski, M.}, \bibinfo{author}{Zeilinger, A.},
  \bibinfo{author}{Horne, M.~A.} \& \bibinfo{author}{Ekert, A.~K.}
\newblock \bibinfo{title}{‘‘event-ready-detectors’’ bell experiment via
  entanglement swapping}.
\newblock \emph{\bibinfo{journal}{Physical Review Letters}}
  \textbf{\bibinfo{volume}{71}}, \bibinfo{pages}{4287} (\bibinfo{year}{1993}).

\bibitem{humphreys2018deterministic}
\bibinfo{author}{Humphreys, P.~C.} \emph{et~al.}
\newblock \bibinfo{title}{Deterministic delivery of remote entanglement on a
  quantum network}.
\newblock \emph{\bibinfo{journal}{Nature}} \textbf{\bibinfo{volume}{558}},
  \bibinfo{pages}{268--273} (\bibinfo{year}{2018}).

\bibitem{TSB+20}
\bibinfo{author}{Tan, E. Y.-Z.} \emph{et~al.}
\newblock \bibinfo{title}{{Improved DIQKD protocols with finite-size
  analysis}}.
\newblock \emph{\bibinfo{journal}{arXiv:2012.08714v1 [quant-ph]}}
  (\bibinfo{year}{2020}).
\newblock \urlprefix\url{https://arxiv.org/abs/2012.08714v1}.

\bibitem{eat}
\bibinfo{author}{Dupuis, F.}, \bibinfo{author}{Fawzi, O.} \&
  \bibinfo{author}{Renner, R.}
\newblock \bibinfo{title}{Entropy accumulation}.
\newblock \emph{\bibinfo{journal}{arXiv:1607.01796}}  (\bibinfo{year}{2016}).

\bibitem{eatqkd}
\bibinfo{author}{Arnon-Friedman, R.}, \bibinfo{author}{Dupuis, F.},
  \bibinfo{author}{Fawzi, O.}, \bibinfo{author}{Renner, R.} \&
  \bibinfo{author}{Vidick, T.}
\newblock \bibinfo{title}{{Practical device-independent quantum cryptography
  via entropy accumulation}}.
\newblock \emph{\bibinfo{journal}{Nature Communications}}
  \textbf{\bibinfo{volume}{9}}, \bibinfo{pages}{459} (\bibinfo{year}{2018}).

\bibitem{eatsecondorder}
\bibinfo{author}{Dupuis, F.} \& \bibinfo{author}{Fawzi, O.}
\newblock \bibinfo{title}{{Entropy Accumulation with Improved Second-Order
  Term}}.
\newblock \emph{\bibinfo{journal}{IEEE Transactions on Information Theory}}
  \textbf{\bibinfo{volume}{65}}, \bibinfo{pages}{7596--7612}
  (\bibinfo{year}{2019}).

\bibitem{Zhang2020Efficient}
\bibinfo{author}{Zhang, Y.}, \bibinfo{author}{Fu, H.} \&
  \bibinfo{author}{Knill, E.}
\newblock \bibinfo{title}{Efficient randomness certification by quantum
  probability estimation}.
\newblock \emph{\bibinfo{journal}{Physical Review Research}}
  \textbf{\bibinfo{volume}{2}}, \bibinfo{pages}{013016} (\bibinfo{year}{2020}).
\newblock
  \urlprefix\url{https://link.aps.org/doi/10.1103/PhysRevResearch.2.013016}.

\bibitem{ourworkarxiv}
\bibinfo{author}{Tan, E. Y.-Z.}, \bibinfo{author}{Schwonnek, R.},
  \bibinfo{author}{Goh, K.~T.}, \bibinfo{author}{Primaatmaja, I.~W.} \&
  \bibinfo{author}{Lim, C. C.-W.}
\newblock \bibinfo{title}{{Computing secure key rates for quantum key
  distribution with untrusted devices}}.
\newblock \emph{\bibinfo{journal}{arXiv:1908.11372v1}}  (\bibinfo{year}{2019}).

\bibitem{devetak}
\bibinfo{author}{Devetak, I.} \& \bibinfo{author}{Winter, A.}
\newblock \bibinfo{title}{Distillation of secret key and entanglement from
  quantum states}.
\newblock \emph{\bibinfo{journal}{Proceedings of the Royal Society A —
  Mathematical, Physical and Engineering Sciences}}
  \textbf{\bibinfo{volume}{461}}, \bibinfo{pages}{207--235}
  (\bibinfo{year}{2005}).

\bibitem{liebfrank}
\bibinfo{author}{Frank, R.~L.} \& \bibinfo{author}{Lieb, E.~H.}
\newblock \bibinfo{title}{Entropy and the uncertainty principle}.
\newblock \emph{\bibinfo{journal}{Annales Henri Poincar{\'e}}}
  \textbf{\bibinfo{volume}{13}}, \bibinfo{pages}{1711--1717}
  (\bibinfo{year}{2012}).
\newblock \urlprefix\url{https://doi.org/10.1007/s00023-012-0175-y}.

\bibitem{junge}
\bibinfo{author}{Gao, L.}, \bibinfo{author}{Junge, M.} \&
  \bibinfo{author}{LaRacuente, N.}
\newblock \bibinfo{title}{Uncertainty principle for quantum channels}.
\newblock In \emph{\bibinfo{booktitle}{2018 IEEE International Symposium on
  Information Theory (ISIT)}}, \bibinfo{pages}{996--1000}
  (\bibinfo{year}{2018}).

\bibitem{eureview1}
\bibinfo{author}{Wehner, S.} \& \bibinfo{author}{Winter, A.}
\newblock \bibinfo{title}{Entropic uncertainty relations {--} a survey}.
\newblock \emph{\bibinfo{journal}{New Journal of Physics}}
  \textbf{\bibinfo{volume}{12}}, \bibinfo{pages}{025009}
  (\bibinfo{year}{2010}).
\newblock
  \bibinfo{note}{\href{https://arxiv.org/abs/0907.3704}{{arXiv}:0907.3704}}.

\bibitem{thesis}
\bibinfo{author}{Schwonnek, R.}
\newblock \bibinfo{title}{Uncertainty relations in quantum theory}
  (\bibinfo{year}{2018}).
\newblock \bibinfo{note}{Ph.D thesis, Leibniz University Hannover,
  https://doi.org/10.15488/3600}.

\bibitem{liu}
\bibinfo{author}{Liu, S.}, \bibinfo{author}{Mu, L.-Z.} \& \bibinfo{author}{Fan,
  H.}
\newblock \bibinfo{title}{Entropic uncertainty relations for multiple
  measurements}.
\newblock \emph{\bibinfo{journal}{Physical Review A}}
  \textbf{\bibinfo{volume}{91}}, \bibinfo{pages}{042133}
  (\bibinfo{year}{2015}).

\bibitem{deutsch}
\bibinfo{author}{Deutsch, D.}
\newblock \bibinfo{title}{Uncertainty in quantum measurements}.
\newblock \emph{\bibinfo{journal}{Physical Review Letters}}
  \textbf{\bibinfo{volume}{50}}, \bibinfo{pages}{631} (\bibinfo{year}{1983}).

\bibitem{hans2}
\bibinfo{author}{Maassen, H.}
\newblock \emph{\bibinfo{title}{Discrete entropic uncertainty relation}}
  (\bibinfo{publisher}{Springer}, \bibinfo{year}{1990}).
\newblock \bibinfo{note}{`Quantum Probability and Applications V' (Proceedings
  {H}eidelberg 1988),Lecture Notes in Mathematics 1442}.

\bibitem{muff}
\bibinfo{author}{Maassen, H.} \& \bibinfo{author}{Uffink, J.~B.}
\newblock \bibinfo{title}{Generalized entropic uncertainty relations}.
\newblock \emph{\bibinfo{journal}{Physical Review Letters}}
  \textbf{\bibinfo{volume}{60}}, \bibinfo{pages}{1103} (\bibinfo{year}{1988}).

\bibitem{ourentro}
\bibinfo{author}{Abdelkhalek, K.} \emph{et~al.}
\newblock \bibinfo{title}{Optimality of entropic uncertainty relations}.
\newblock \emph{\bibinfo{journal}{Int.\ J.\ Quant.\ Inf.}}
  \textbf{\bibinfo{volume}{13}}, \bibinfo{pages}{1550045}
  (\bibinfo{year}{2015}).
\newblock
  \bibinfo{note}{\href{https://arxiv.org/abs/1509.00398}{{arXiv}:1509.00398}}.

\bibitem{schneeloch}
\bibinfo{author}{Schneeloch, J.}, \bibinfo{author}{Broadbent, C.~J.},
  \bibinfo{author}{Walborn, S.~P.}, \bibinfo{author}{Cavalcanti, E.~G.} \&
  \bibinfo{author}{Howell, J.~C.}
\newblock \bibinfo{title}{{Einstein-Podolsky-Rosen} steering inequalities from
  entropic uncertainty relations}.
\newblock \emph{\bibinfo{journal}{Physical Review A}}
  \textbf{\bibinfo{volume}{87}}, \bibinfo{pages}{062103}
  (\bibinfo{year}{2013}).
\newblock
  \bibinfo{note}{\href{https://arxiv.org/abs/1303.7432}{{arXiv}:1303.7432}}.

\bibitem{additivity}
\bibinfo{author}{Schwonnek, R.}
\newblock \bibinfo{title}{Additivity of entropic uncertainty relations}.
\newblock \emph{\bibinfo{journal}{Quantum}} \textbf{\bibinfo{volume}{2}},
  \bibinfo{pages}{59} (\bibinfo{year}{2018}).

\bibitem{alberto1}
\bibinfo{author}{Riccardi, A.}, \bibinfo{author}{Macchiavello, C.} \&
  \bibinfo{author}{Maccone, L.}
\newblock \bibinfo{title}{Tight entropic uncertainty relations for systems with
  dimension three to five}.
\newblock \emph{\bibinfo{journal}{Physical Review A}}
  \textbf{\bibinfo{volume}{95}}, \bibinfo{pages}{032109}
  (\bibinfo{year}{2017}).
\newblock
  \bibinfo{note}{\href{https://arxiv.org/abs/1701.04304}{{arXiv}:1701.04304}}.

\bibitem{Tomamichel2013}
\bibinfo{author}{Tomamichel, M.} \& \bibinfo{author}{Hänggi, E.}
\newblock \bibinfo{title}{The link between entropic uncertainty and
  nonlocality}.
\newblock \emph{\bibinfo{journal}{Journal of Physics A: Mathematical and
  Theoretical}} \textbf{\bibinfo{volume}{46}}, \bibinfo{pages}{055301}
  (\bibinfo{year}{2013}).
\newblock
  \urlprefix\url{https://doi.org/10.1088%2F1751-8113%2F46%2F5%2F055301}.

\bibitem{amu}
\bibinfo{author}{Dammeier, L.}, \bibinfo{author}{Schwonnek, R.} \&
  \bibinfo{author}{Werner, R.}
\newblock \bibinfo{title}{Uncertainty relations for angular momentum}.
\newblock \emph{\bibinfo{journal}{New Journal of Physics}}
  \textbf{\bibinfo{volume}{9}}, \bibinfo{pages}{093946} (\bibinfo{year}{2015}).
\newblock
  \bibinfo{note}{\href{https://arxiv.org/abs/1505.00049}{{arXiv}:1505.00049}}.

\bibitem{bottcher}
\bibinfo{author}{Böttcher, A.} \& \bibinfo{author}{Spitkovsky, I.}
\newblock \bibinfo{title}{A gentle guide to the basics of two projections
  theory}.
\newblock \emph{\bibinfo{journal}{Linear Algebra and its Applications}}
  \textbf{\bibinfo{volume}{432}}, \bibinfo{pages}{1412 -- 1459}
  (\bibinfo{year}{2010}).

\bibitem{halmos}
\bibinfo{author}{Halmos, P.~R.}
\newblock \bibinfo{title}{Two subspaces}.
\newblock \emph{\bibinfo{journal}{Transactions of the American Mathematical
  Society}} \textbf{\bibinfo{volume}{144}}, \bibinfo{pages}{381--389}
  (\bibinfo{year}{1969}).

\bibitem{boyd04p8}
\bibinfo{author}{Boyd, S.} \& \bibinfo{author}{Vandenberghe, L.}
\newblock \emph{\bibinfo{title}{{Convex Optimization}}}
  (\bibinfo{publisher}{Cambridge University Press}, \bibinfo{year}{2004}).

\bibitem{varucr1}
\bibinfo{author}{Schwonnek, R.}, \bibinfo{author}{Dammeier, L.} \&
  \bibinfo{author}{Werner, R.~F.}
\newblock \bibinfo{title}{State-independent uncertainty relations and
  entanglement detection in noisy systems}.
\newblock \emph{\bibinfo{journal}{Physical Review Letters}}
  \textbf{\bibinfo{volume}{119}}, \bibinfo{pages}{170404}
  (\bibinfo{year}{2017}).

\bibitem{varucr2}
\bibinfo{author}{Zhao, Y.-Y.} \emph{et~al.}
\newblock \bibinfo{title}{Entanglement detection by violations of noisy
  uncertainty relations: A proof of principle}.
\newblock \emph{\bibinfo{journal}{Physical Review Letters}}
  \textbf{\bibinfo{volume}{122}}, \bibinfo{pages}{220401}
  (\bibinfo{year}{2019}).
\end{thebibliography}
\end{document}